\documentclass{sn-jnl}

\usepackage[dvipsnames]{xcolor}
\usepackage{graphicx}
\usepackage{subcaption}

\usepackage[T1]{fontenc}
\usepackage{hyperref}

\usepackage[numbers, sort&compress]{natbib}
\setcitestyle{super,open={},close={}}
\usepackage{amsmath}
\usepackage{amsfonts}
\usepackage{graphbox}
\usepackage{multibib}
\newcites{met}{References}
\usepackage[colorinlistoftodos]{todonotes}
\usepackage{listings}
\usepackage{multirow}
\usepackage{array}
\usepackage{tikz}
\definecolor{orange}{HTML}{d9944f}
\usetikzlibrary{patterns,arrows,positioning,fit}
\usetikzlibrary{decorations.pathreplacing}
\usetikzlibrary{decorations.pathmorphing}
\usepackage[linesnumbered,ruled,vlined]{algorithm2e}
\usepackage[capitalize]{cleveref}
\usepackage{booktabs}
\usepackage{adjustbox}      
\usepackage{float}          
\let\orcidlogo\relax  
\usepackage{orcidlink}
\usepackage{lineno}
\usepackage{pdflscape}
\usepackage{makecell}
\usepackage{parskip} 

\usepackage{changepage}

\usepackage{todonotes}
\newboolean{SHOWCOMMENTS}
\setboolean{SHOWCOMMENTS}{true} 
\ifthenelse{\boolean{SHOWCOMMENTS}}{%
  \newcommand{\note}[1]{\colorbox{gray!20}{\parbox{1\linewidth}{##1}}}%
}{%
  \newcommand{\note}[1]{}%
}

\newcommand{\defineComment}[4]{%
  \ifthenelse{\boolean{SHOWCOMMENTS}}{%
    \newcommand{#1}[1]{\todo[color=#3,inline]{\footnotesize{\bfseries #4:} ##1}}%
  }{%
    \newcommand{#1}[1]{}%
  }%
  \newcommand{#2}[1]{{\color{#3}##1}}%
}

\defineComment{\tdMK}{\MK}{black}{Martin}
\defineComment{\tdSK}{\SK}{SkyBlue}{Sascha}
\defineComment{\tdJB}{\JB}{green!90!gray!10}{Julia}
\defineComment{\tdLP}{\LP}{violet!50}{Lena}
\defineComment{\tdCG}{\CG}{SpringGreen!70}{Carlotta}
\defineComment{\tdRS}{\RS}{teal!20}{René}
\defineComment{\tdAW}{\AW}{SeaGreen!50}{Anna}
\defineComment{\tdHT}{\HT}{olive}{Hannah}
\defineComment{\tdHZ}{\HZ}{yellow!80!gray!25}{Henrik}
\defineComment{\tdAS}{\AS}{orange!80!red!25}{Agatha}
\defineComment{\tdDK}{\DK}{red!80!gray!25}{David}
\defineComment{\tdMB}{\MB}{JungleGreen!80}{Max}
\defineComment{\tdDA}{\DA}{red!30!gray}{Daniel A.}
\defineComment{\tdNW}{\NW}{MidnightBlue}{Nils}
\defineComment{\tdPL}{\PL}{gray}{Patrick}
\defineComment{\tdKN}{\KN}{Cyan}{Khoa}

\renewcommand{\d}[1]{\mathrm{d}#1}

\title{MEmilio -- A high performance Modular EpideMIcs simuLatIOn software for multi-scale and comparative simulations of infectious disease dynamics}

\author[1]{\fnm{Julia} \sur{Bicker}\orcidlink{0000-0001-9382-4209}} 
\equalcont{}

\author[2,3]{\fnm{Carlotta} \sur{Gerstein}\orcidlink{0009-0004-4410-0502}}
\equalcont{}

\author[4]{\fnm{David} \sur{Kerkmann}\orcidlink{0009-0007-9109-096X}}
\equalcont{}

\author[1,5]{\fnm{Sascha} \sur{Korf}\orcidlink{0000-0002-1431-3046}}
\equalcont{}

\author[1]{\fnm{Ren\'e} \sur{Schmieding} \orcidlink{0000-0002-2769-0270}}
\equalcont{}

\author[1]{\fnm{Anna} \sur{Wendler}\orcidlink{0000-0002-1816-8907}}
\equalcont{}

\author[1]{\fnm{Henrik} \sur{Zunker}\orcidlink{0000-0002-9825-365X}}
\equalcont{Shared first author in alphabetical order. These authors contributed equally to this work.\\$^+$These authors contributed equally.} 

\author[1]{\fnm{Daniel} \sur{Abele}} 

\author[1,7]{\fnm{Maximilian} \sur{Betz}\orcidlink{0009-0001-9181-9999}}

\author[6]{\fnm{Khoa} \sur{Nguyen}\orcidlink{0000-0002-3123-602X}} 

\author[1]{\fnm{Lena} \sur{Pl\"otzke}\orcidlink{0000-0003-0440-1429}} 

\author[2,3]{\fnm{Kilian} \sur{Volmer}\orcidlink{0009-0008-9861-0216}}

\author[1]{\fnm{Agatha} \sur{Schmidt}\orcidlink{0009-0006-5766-8804}}

\author[1,2,3]{\fnm{Nils} \sur{Wa\ss muth}}

\author[1]{\fnm{Patrick} \sur{Lenz}\orcidlink{0009-0000-1535-0626}}

\author[1]{\fnm{Daniel} \sur{Richter}}  

\author[1,8]{\fnm{Hannah} \sur{Tritzschak}\orcidlink{0009-0003-8543-662X}} 

\author[7]{\fnm{Ralf} \sur{Hannemann-Tamas}}

\author[7]{\fnm{Julian} \sur{Litz}} 

\author[1,9]{\fnm{Paul} \sur{Johannssen}\orcidlink{0009-0006-7869-7647}} 

\author[1]{\fnm{Marielena} \sur{Borges}}  

\author[1]{\fnm{Annika} \sur{Jungklaus}} 

\author[1]{\fnm{Manuel} \sur{Heger}} 

\author[1]{\fnm{Annalena} \sur{Lange}} 

\author[1]{\fnm{Elisabeth} \sur{Kluth}} 

\author[1]{\fnm{Kathrin} \sur{Rack}\orcidlink{0000-0002-5794-5705}} 

\author[2,3]{\fnm{Vincent} \sur{Wieland}\orcidlink{0009-0003-9592-8871}}

\author[2,3]{\fnm{Jonas} \sur{Arruda}\orcidlink{0009-0008-9644-5771}}

\author[4]{\fnm{Sebastian C.} \sur{Binder}\orcidlink{0000-0003-1169-1786}} 

\author[10]{\fnm{Margrit} \sur{Klitz}\orcidlink{0000-0003-3657-4180}} 

\author[11]{\fnm{Martin} \sur{Siggel}} 

\author[7]{\fnm{Manuel} \sur{Dahmen}\orcidlink{0000-0003-2757-5253}} 

\author[1]{\fnm{Achim} \sur{Basermann}\orcidlink{0000-0003-3637-3231}} 

\author[4,12,13,+]{\fnm{Michael} \sur{Meyer-Hermann}\orcidlink{0000-0002-4300-2474}}

\author[2,3,+]{\fnm{Jan} \sur{Hasenauer}\orcidlink{0000-0002-4935-3312}}

\author*[1,2,3]{\fnm{Martin J.} \sur{K\"uhn}\orcidlink{0000-0002-0906-6984}}
\email{martin.kuehn@dlr.de}

\affil*[1]{\orgdiv{Institute of Software Technology, Department of High-Performance Computing}, \orgname{German Aerospace Center}, \orgaddress{\city{Cologne}, \postcode{51147}, \country{Germany}}}

\affil[2]{\orgdiv{Bonn Center for Mathematical Life Sciences} \orgname{University of Bonn}, \orgaddress{\city{Bonn}, \postcode{53115}, \country{Germany}}}

\affil[3]{\orgdiv{Life and Medical Sciences Institute} \orgname{University of Bonn}, \orgaddress{\city{Bonn}, \postcode{53115}, \country{Germany}}}

\affil[4]{\orgdiv{Department of Systems Immunology and Braunschweig Integrated Centre of Systems Biology}, \orgname{Helmholtz Centre for Infection Research}, \orgaddress{\city{Brunswick}, \postcode{38124}, \country{Germany}}}

\affil[5]{\orgdiv{Institute of Bio- and Geosciences: Biotechnology (IBG-1)}, \orgname{Forschungszentrum Jülich GmbH}, \orgaddress{\city{Jülich}, \postcode{52425}, \country{Germany}}}

\affil[6]{\orgname{Centre universitaire de m\'edecine g\'en\'erale et sant\'e publique - Unisant\'e}, \orgaddress{\city{Lausanne}, \postcode{1010}, \country{Switzerland}}}

\affil[7]{\orgdiv{Institute of Climate and Energy Systems: Energy Systems Engineering (ICE-1)}, \orgname{Forschungszentrum Jülich GmbH}, \orgaddress{\city{Jülich}, \postcode{52425}, \country{Germany}}}

\affil[8]{\orgdiv{Institute for Numerical Simulation}, \orgname{University of Bonn}, \orgaddress{\city{Bonn}, \postcode{53115}, \country{Germany}}}

\affil[9]{\orgdiv{Translational Neuroimaging, Department for Neuroradiology} \orgname{Universitätsklinikum Bonn}, \orgaddress{\city{Bonn}, \postcode{53127}, \country{Germany}}}

\affil[10]{ \orgdiv{German Space Agency at DLR}, \orgname{German Aerospace Center}, \orgaddress{\city{Cologne}, \postcode{51147}, \country{Germany}}}

\affil[11]{\orgdiv{Institute of Propulsion Technology, Engine Department}, \orgname{German Aerospace Center}, \orgaddress{\city{Cologne}, \postcode{51147}, \country{Germany}}}

\affil[12]{\orgdiv{Institute for Biochemistry, Biotechnology and Bioinformatics} \orgname{Technische Universität Braunschweig}, \orgaddress{\city{Braunschweig}, \country{Germany}}}
\affil[13]{\orgname{Lower Saxony Center for Artificial Intelligence and Causal Methods in Medicine (CAIMed)}, \orgaddress{\city{Hannover}, \country{Germany}}}
       
\date{January 2026}

\abstract{
Epidemic and pandemic preparedness with rapid outbreak response rely on timely, trustworthy evidence. Mathematical models are crucial for supporting timely and reliable evidence generation for public health decision-making with models spanning approaches from compartmental and metapopulation models to detailed agent-based simulations. Yet, the accompanying software ecosystem remains fragmented across model types, spatial resolutions, and computational targets, making models harder to compare, extend, and deploy at scale. Here we present MEmilio, a modular, high-performance framework for epidemic simulation that harmonizes the specification and execution of diverse dynamic epidemiological models within a unified and harmonized architecture. MEmilio couples an efficient C++ simulation core with coherent model descriptions and a user-friendly Python interface, enabling workflows that run on laptops as well as high-performance computing systems. Standardized representations of space, demography, and mobility support straightforward adaptations in resolution and population size, facilitating systematic inter-model comparisons and ensemble studies. The framework integrates readily with established tools for uncertainty quantification and parameter inference, supporting a broad range of applications from scenario exploration to calibration. Finally, strict software-engineering practices, including extensive unit and continuous integration testing, promote robustness and minimize the risk of errors as the framework evolves. By unifying implementations across modeling paradigms, MEmilio aims to lower barriers to reuse and generalize models, enable principled comparisons of implicit assumptions, and accelerate the development of novel approaches that strengthen modeling-based outbreak preparedness.
}

\keywords{Infectious diseases, public health action, interventions, agent-based modeling, metapopulation modeling, high-performance computing}

\begin{document}

\maketitle

Pandemic preparedness and rapid outbreak response are increasingly important in a globalized world with a rising frequency of epidemics and pandemics~\cite{daszak_workshop_2020}. In this context, timely and reliable evidence generation is essential to support public health decision-making. Over the past decades, mathematical modeling of infectious disease dynamics has become a central tool for hypothesis generation and policy evaluation, and during the COVID-19 pandemic public institutions such as the European Centre for Disease Prevention and Control highlighted mathematical modeling as one of the principal sources of evidence for assessing the effectiveness of interventions~\cite{ecdc_npis_2024}. These findings, together with the continued circulation of seasonal respiratory pathogens such as influenza and RSV, as well as re-emerging threats including Ebola and Dengue, underscore the need for sustained modeling capacity that can be rapidly mobilized to provide robust, evidence-informed support during outbreaks.

From a methodological perspective, infectious disease modeling spans a spectrum of formulations that differ in granularity, data requirements, computational cost, and interpretability. Population-based compartmental models (PBMs) describe aggregated disease dynamics using ordinary~\cite{brauer_mathematical_2019,epidemik,grunnill_metacast_2024,adhikari2020pyross}, stochastic~\cite{adhikari2020pyross,epydemix,andersson_stochastic_2012}, or integro-differential equations~\cite{brauer_age_2009,messina_non-standard_2022}. Metapopulation models (MPMs)~\cite{pei_differential_2020,kuhn_assessment_2021,chen_compliance_2021} extend these approaches by incorporating spatial structure and coupling regions through mobility, thus enabling the study of spatio-temporal heterogeneity and regionally targeted nonpharmaceutical interventions (NPIs). Even further, agent-based models (ABMs)~\cite{bicher_agent-based_2013,grefenstette_fred_2013,collier_parallel_2013,OpenABM,bershteyn_implementation_2018,kerr_covasim_2021,mueller2021episim,shattock_impact_2022,KERKMANN2025110269,chen_epihiperhigh_2025,Ponge23} resolve individuals and their interactions, allowing explicit representations of individual contact patterns, mobility, and behavior. Hybrid approaches in time and space~\cite{adhikari2020pyross,bradhurst_hybrid_2015,bicker_hybrid_2025,bostanci2025,kehrer2025hybridabmpdeframeworkrealworld} aim to combine these paradigms in order to balance realism and computational feasibility. However, a single model formulation can barely address all facets of infectious disease mitigation and the research question typically determines which model type is most appropriate.

\begin{figure}[!t]
    \begin{adjustwidth}{0.0in}{0in}
    \centering
    \includegraphics[width=0.97\linewidth]{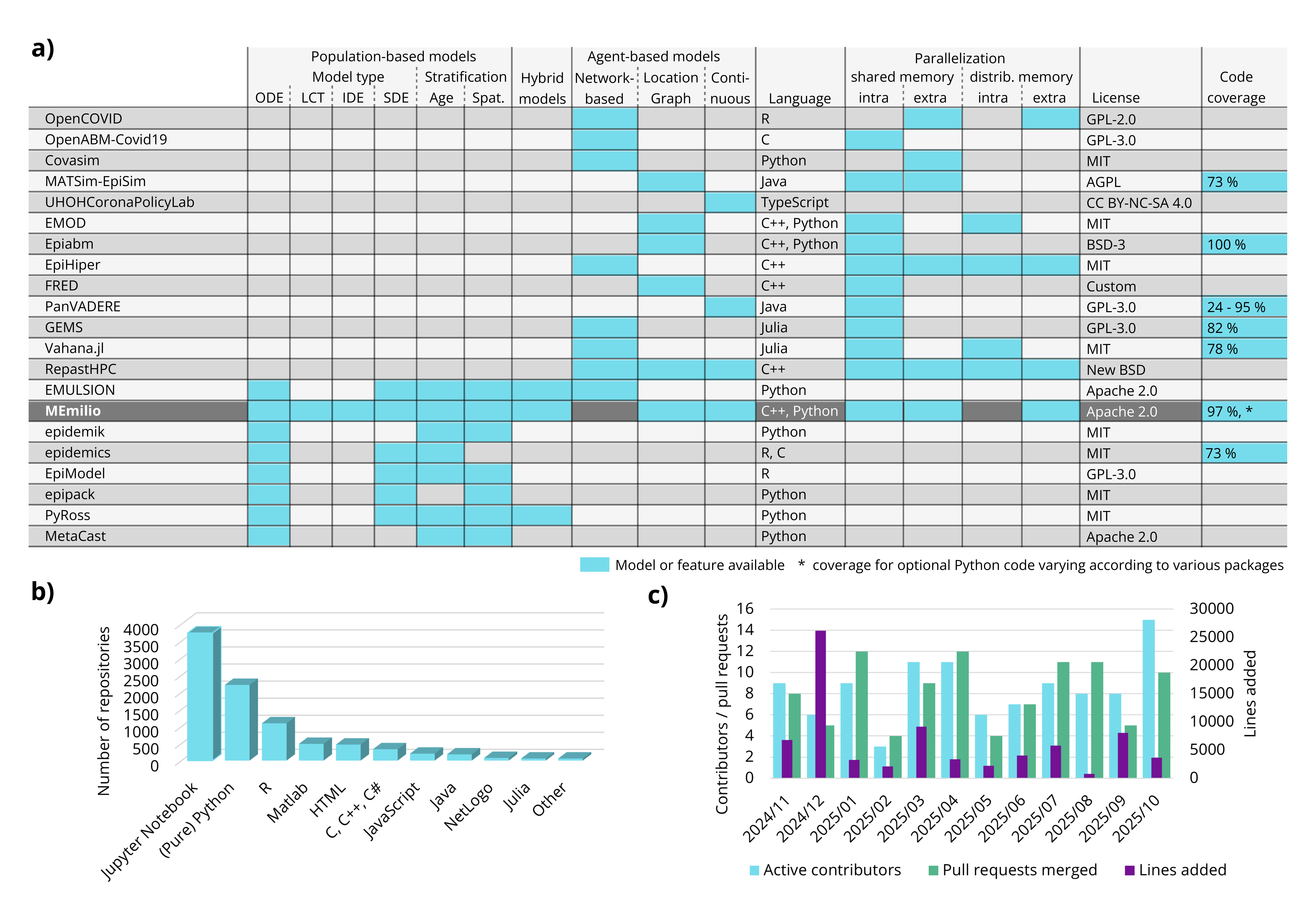}
    \caption{\textbf{Overview of open-source, state-of-the-art infectious disease modeling frameworks, highlighting the distinctions of MEmilio.} \textbf{a)} Overview of 21 software frameworks, depicting which model types are available through the software, which programming language has been used, if parallelization techniques are provided, which license is given, and, if unit and software tests are implemented, which code coverage is achieved. \textbf{b)} Statistical overview of the 10 most used programming languages in public GitHub repositories according to the search terms: {SARS-CoV-2, COVID-19, influenza, Ebola, HIV, malaria, transmission, epidemiological, infection, pandemic, epidemic, endemic, pandemic spread, epidemic spread, endemic spread} (accessed through the GitHub API). \textbf{c)} Overview of active development of MEmilio, expanding for new applications and features.}
    \label{fig:model_comparisons}
    \end{adjustwidth}
\end{figure}

A wide ecosystem of open-source software supports infectious disease modeling across this spectrum, typically specializing in a particular model class and resolution (\cref{fig:model_comparisons}a). While PBMs are often reimplemented for particular applications, different packages such as epidemik~\cite{epidemik}, MetaCast~\cite{grunnill_metacast_2024}, PyRoss~\cite{adhikari2020pyross}, Epydemix~\cite{epydemix}, epipack~\cite{epipack}, EpiModel~\cite{jenness_epimodel_2018}, epidemics~\cite{Gupte_epidemics_Composable_Epidemic}, or EMULSION~\cite{picault_emulsion_2019} also allow the flexible modeling through PBMs and MPMs; mostly using ordinary or stochastic differential equations (ODEs or SDEs). For ABMs, a variety of implementations exist. A relevant number of ABMs, such as RepastHPC~\cite{collier_parallel_2013}, OpenABM~\cite{OpenABM}, Covasim~\cite{kerr_covasim_2021}, OpenCOVID~\cite{shattock_impact_2022}, EpiHiper~\cite{chen_epihiperhigh_2025}, GEMS~\cite{Ponge23}, or Vahana.jl~\cite{fuerst2024vahanajlframeworknot}, are implemented as network-based models. These models commonly operate on coarser temporal discretizations (e.g., daily time steps) and represent locations implicitly, while mobility- or activity-based ABMs such as FRED~\cite{grefenstette_fred_2013}, EMOD~\cite{bershteyn_implementation_2018}, MatSim-EpiSim~\cite{mueller2021episim}, UHOHCoronaPolicyLab~\cite{vermeulen2021}, PanVADERE~\cite{rahn_modelling_2022}, or the models by Goldebogen et al.~\cite{Goldebogen22} and Cuevas~\cite{CUEVAS2020103827} explicitly simulate movement and allow explicit modeling inside locations, typically at a more fine-grained temporal resolution. 

Despite this breadth, the current software landscape remains fragmented across model classes, levels of resolution, and computational orientation; see~\cref{fig:model_comparisons}a for an overview \MK{and comparison of MEmilio with} 20 relevant open-source software frameworks or packages \MK{that have been selected after searching on GitHub and in the corresponding scientific literature}. In particular, compartmental modeling software often lacks high-performance computing (HPC) support for large-scale parameter estimation and uncertainty analysis. In addition, it offers limited flexibility for data-driven, nonexponential state transition distributions and is mostly not connected to finer-granular or hybrid modeling approaches. While several tools provide efficient simulation capabilities, integrated workflows that combine scalable calibration, uncertainty quantification, and sensitivity analysis with modern epidemiological data pipelines remain uncommon. Conversely, detailed agent-based frameworks are rarely linked to corresponding coarser or hybrid models that would allow for rapid early exploration of open research questions before committing to computationally demanding simulations; many frameworks also have limited scaling capabilities with respect to population sizes and HPC infrastructure. More broadly, the absence of harmonized abstractions across modeling paradigms complicates systematic inter-model comparison: even when similar epidemiological processes are represented, models are typically implemented in tool-specific ways that hinder the transparent transfer of assumptions, intervention definitions, and data interfaces across PBMs, MPMs, and ABMs. Eventually, there is currently no open framework that allows different epidemiological model types -- ranging from PBMs to MPMs and ABMs -- to be generated, calibrated, and compared from a shared abstract description of the relevant processes, while additionally aiming at computational scalability and software-engineering robustness.

\begin{figure}[!b]
    \begin{adjustwidth}{-0.in}{0in}
    \centering
    \includegraphics[width=0.98\linewidth]{MEmilio_figures_2.pdf}
    \caption{\textbf{Overview of the MEmilio software framework.} The MEmilio software framework consists of an efficient C++ backend (center) which provides models of different types, granularity and application foci. Several optional modules (left and right) provide additional functionality to be used from MEmilio's C++ or Python interface. \MK{Software packages located with MEmilio's Python frontend are shown in green, additional functionality directly integrated in the C++ backend is shown in blue. C++, Python, and SBML logos in the upper part of the boxes indicate the (main) programming language of the additional functionality or package.} In the bottom row, we depict the execution on consumer laptops or supercomputing infrastructure, together with the typical external packages that can be integrated into a modeling workflow.}
    \label{fig:MEmilio_overview}
    \end{adjustwidth}
\end{figure}

To address these challenges, we present the \textit{MEmilio} high performance Modular EpideMIcs simuLatIOn software framework for simulating infectious disease dynamics, developed as a unified and scalable framework. Distinct features of MEmilio include
\begin{itemize}
  \setlength{\topsep}{2pt}     
  \setlength{\parsep}{0pt}     
  \setlength{\partopsep}{0pt}  
    \item \textbf{uniform model descriptions} and \textbf{harmonized implementations} of a broad range of epidemiological model types from \textbf{PBMs to MPMs and ABMs};
    \item \textbf{efficient C++ backends} with coherent interfaces and a \textbf{user-friendly Python frontend}~\cite{BR25} supporting execution from \textbf{laptops and supercomputing infrastructure} as well as the use of established Python packages for uncertainty quantification and parameter inference;
    \item explicit support for \textbf{multiple compartmental formulations}, including ODE and SDE as well as Linear Chain Trick (LCT) and integro-differential equation (IDE) based models that allow flexible, nonexponential state transition time distributions;
    \item integration of models across scales, including \textbf{hybrid approaches} that facilitate rapid exploration while preserving desired accuracy in focus regimes;
    \item \textbf{standardized spatial and demographic stratifications and mobility layers} enabling scalability in demographic resolution and systematic inter-model comparison;
    \item \textbf{strict software-engineering practices} with extensive and continuous unit and software testing. 
\end{itemize}
MEmilio's backend is extended by two packages enabling \textbf{optimal control}~\cite{betts2010practical} for PBMs and MPMs and by supporting the \textbf{import of models} specified in the Systems Biology Markup Language (SBML) standard~\cite{hucka_systems_2003}. 
MEmilio's machine-learning surrogate model package support Multi-Layer Perceptrons, Long Short-Term Memory networks~\cite{lstm}, Convolutional Neural Networks~\cite{li_survey_2022}, and spatially resolved Graph Neural Networks (GNNs)~\cite{wu_comprehensive_2021} on top of PBMs and MPMs, enabling on-the-fly computation in low-barrier web applications~\cite{betz_esid_2023} \MK{and integration in HPC-capable full software stacks for pandemic mitigation~\cite{loki_infrastructure_2025}}.

\section*{Results}

\subsection*{Applicability to a multitude of research questions and use cases}

\begin{figure}[!b]
    \begin{adjustwidth}{0in}{0in}
    \centering
    \includegraphics[width=0.98\linewidth]{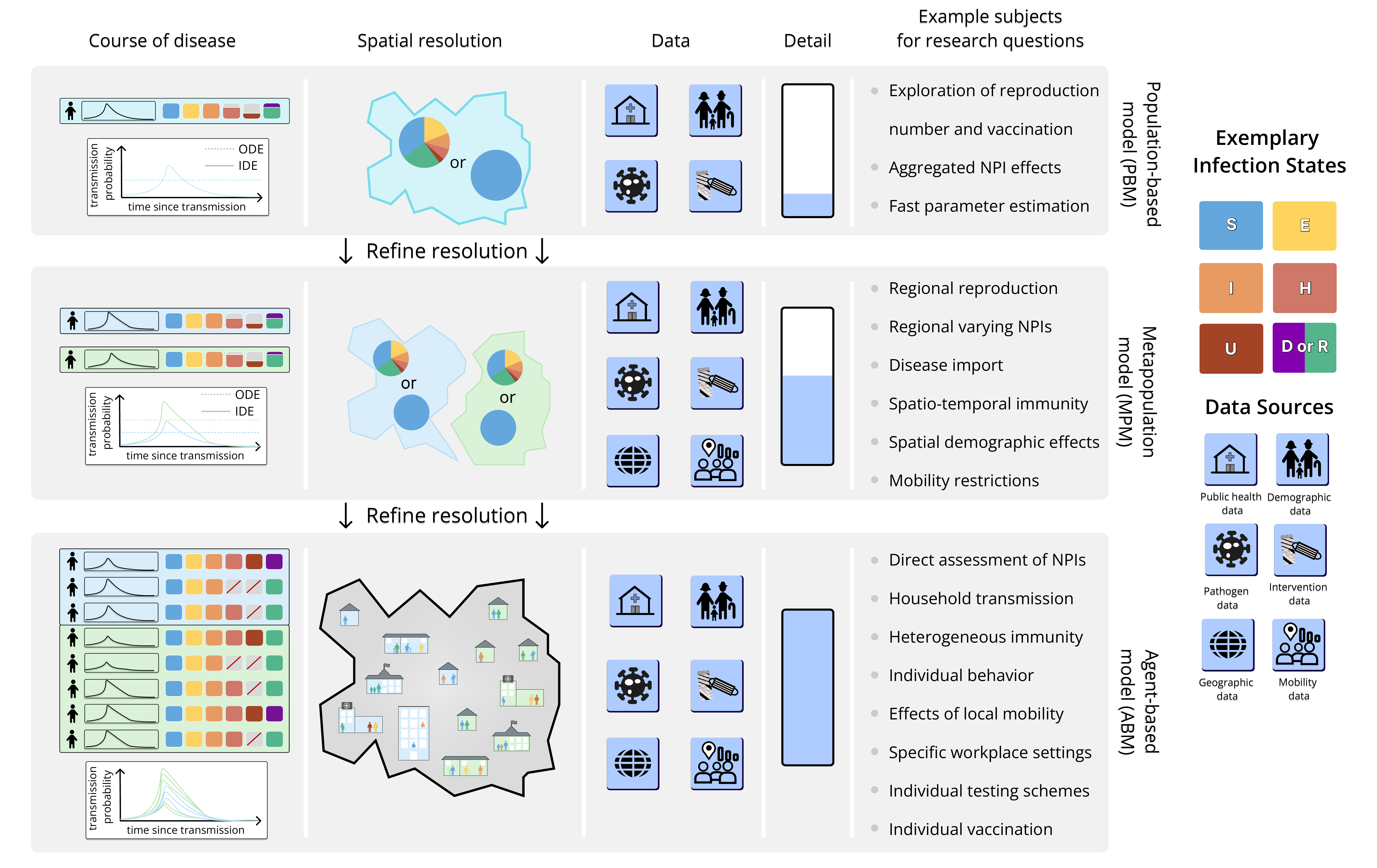}
    \caption{\textbf{Detailed overview of PBMs, MPMs, and ABMs in MEmilio.} Models are shown from finest (top) to coarsest (bottom) granularity. The first column shows the individual (ABM), regionally heterogeneous (MPM) or aggregated (PBM) disease courses, with the corresponding spatial resolution in the second column. The necessary data items with required level of detail are shown in the third and fourth column. Exemplary research questions are given in the rightmost column.}
    \label{fig:different_models_and_questions}
    \end{adjustwidth}
\end{figure}

To address the aforementioned needs, we developed MEmilio as a unified framework supporting infectious disease modeling across multiple levels of granularity and application domains (\cref{fig:MEmilio_overview}). MEmilio provides a broad range of PBMs, MPMs, and ABMs, each with distinct strengths and limitations and each suited for different classes of research questions (\cref{fig:different_models_and_questions}). All models are implemented with harmonized interfaces, consistent data structures, and shared design principles, enabling systematic comparison and facilitating transitions between model types. \MK{In addition to supporting the individual model classes, this common implementation allows models of different fidelity to be applied to matched epidemic settings. Computationally inexpensive PBMs and MPMs can thereby be used to screen parameters, assumptions, and scenarios before selected configurations are investigated with more detailed ABMs. The resulting cross-fidelity comparisons can identify which assumptions materially affect the outcomes and, consequently, which inputs and regimes should be represented when constructing or validating surrogate models for computationally expensive simulations.} For more details on the explicit models, see the Methods section and the Supplementary Information.

Among all considered model formulations, MEmilio's PBMs (\cref{fig:different_models_and_questions}, top row) have the sparsest data requirements and enable quick exploration of reproduction numbers as well as population-level effects of vaccination strategies. In addition, they can be used to assess aggregated NPI effects and are well-suited for early-stage, coarse-grained parameter estimation. Particular models are ODE-based~\cite{kuhn_assessment_2021}, using the LCT~\cite{ploetzke_lct_2025} or more flexible IDE-based representations~\cite{Wendler2025IDE} which allow for transmission probabilities and state transitions dependent on the time since infection or last transition. More realistic distribution assumptions on relevant parameters such as the incubation period makes MEmilio well-suited for analyzing epidemic change points.

MEmilio's PBMs are provided to users such that a simple \texttt{Model($N_G$)} creates a user-defined model with $N_G$ demographic groups including group-specific parameters and contact matrices (\cref{fig:MEmilio_overview}, ``Demographic stratification'') to, e.g., study age-dependent severity or mortality risks. The contact matrix can be further resolved by a simple list of \texttt{ContactLocation}s (such as Home, School, Work, and Other), automatically preparing the inclusion of any type of contact patterns specific by location such as given by POLYMOD~\cite{mossong_social_2008}, Fumanelli et al.~\cite{fumanelli_inferring_2012}, or CoMix~\cite{JARVIS2024100778}. Most importantly, users can thereby separately target locations for interventions or restrictions. 

MEmilio's structure allows users to couple any number $N_P$ of local PBMs with an $N_P\times N_P$ mobility matrix to form an MPM (\cref{fig:MEmilio_overview}, "Metapopulation \& Mobility") that accounts for spatial heterogeneity. MPMs (\cref{fig:different_models_and_questions}, central row) allow for the study of spatially heterogeneous disease dynamics and the effect of locally varying NPIs~\cite{koslow_appropriate_2022}. They explicitly account for disease import from hotspot regions into susceptible populations and can capture spatio-temporal heterogeneity in population-level immunity, thereby enabling the study of complex and region-specific vaccination strategies. As mobility plays a crucial role in shaping disease dynamics, MEmilio supports to study the effects of mobility restrictions and commuter-based testing strategies~\cite{kuhn_regional_2022}. A particularly important feature are region- and threshold-specific intervention schemes. Through a structure denoted \texttt{DynamicNPIs}, user-defined intervention schemes can be automatically applied upon exceeding region-specific thresholds in, e.g., new symptomatic cases or hospitalizations. In combination with optimal control, this structure drastically reduces the number of control variables that otherwise scale with the spatial and temporal resolution and, thus, enables optimal control even for large models. 

MEmilio's ABM, based on individual activities and mobility, (\cref{fig:different_models_and_questions}, bottom row) implements detailed individual transmission probabilities such that users can study superspreading events~\cite{LI2021107788,korf_effect_2025}
or location-specific transmissions in schools, workplaces and recreation areas. The ABM further enables the direct modeling of mask usage, heterogeneous immunity, and usage of NPIs tailored to individual age groups, locations, or symptoms. Furthermore, MEmilio's ABM allows studying (imperfect) quarantining and isolation strategies of variable length, individual (non)compliance, or complex vaccination patterns tailored to individual immunization histories. 

From the aforementioned software packages and frameworks, EMULSION's~\cite{picault_emulsion_2019} approach comes closest to MEmilio, yet, it does not provide HPC support. MEmilio's parallelization approaches comprise shared memory parallelization inside most costly ABM segments and simulation ensemble parallelization for PBMs, MPMs, and ABMs. Altogether, this makes all of MEmilio's models suitable to be simulated with short runtimes on individual computers as well as highly parallelized on supercomputing infrastructure.

Beyond the plethora of MEmilio's functions, MEmilio is a rapidly expanding framework with approximately 10 monthly merged pull requests (\cref{fig:model_comparisons}c) adding new features, novel models and applications in, e.g., veterinary diseases. Adhering to strict software engineering principles with code review, software testing, and continuous integration, MEmilio ensures software stability with minimized error potential for extensions.

\subsection*{Analysis of different diseases and spatio-temporal dynamics across scales}

\begin{figure}[!t]
    \centering
    \includegraphics[width=0.97\linewidth]{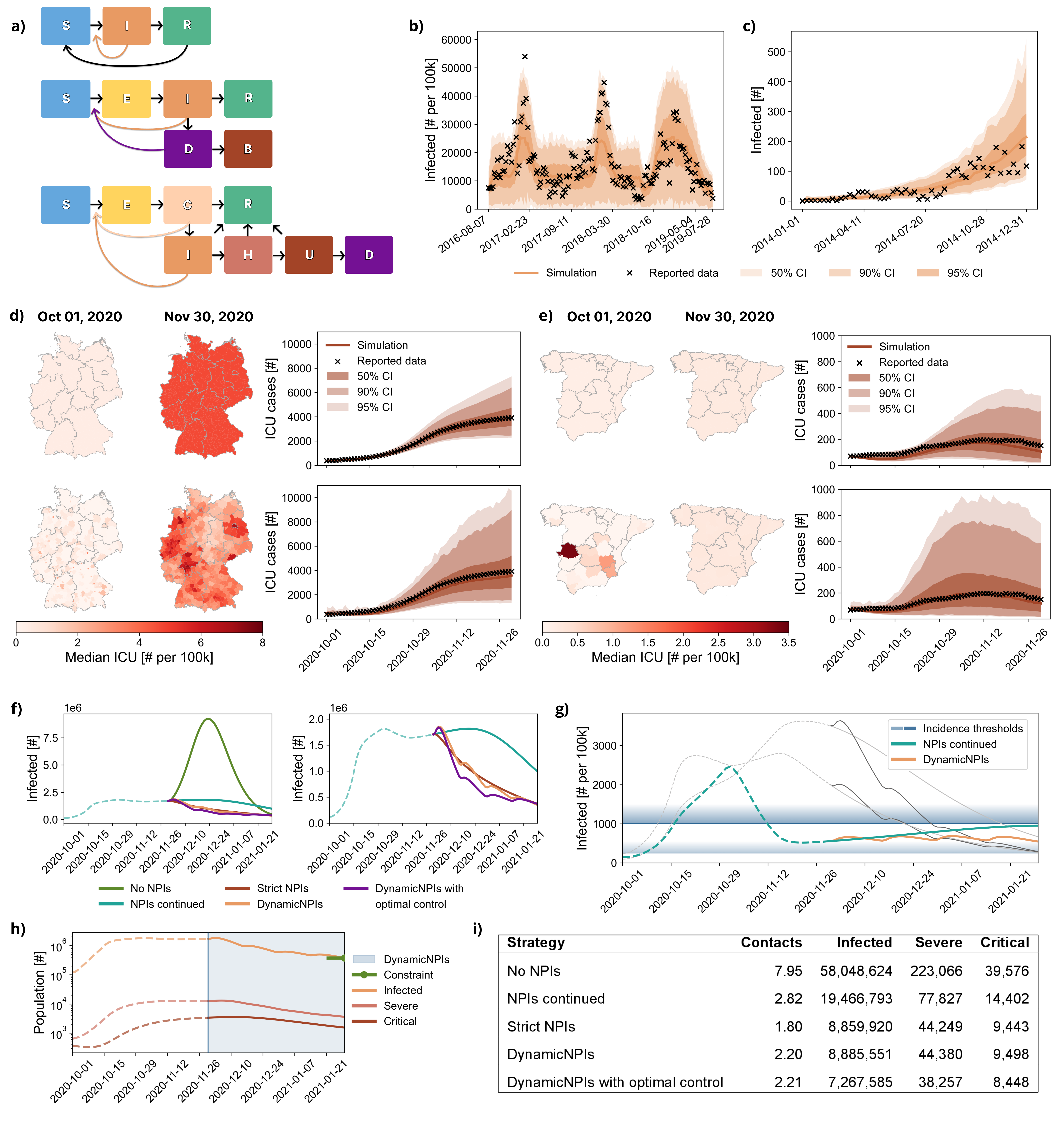}
    \caption{\textbf{Models and calibrations for different diseases and spatio-temporal dynamics together with an evaluation of intervention strategies.} \textbf{a)} SIRS, SEIRDB, and SECIR-type course of disease structure for modeling influenza, Ebola, and early SARS-CoV-2, respectively. \textbf{b)} Fit of the SDE SIRS-type influenza model for Germany. \textbf{c)} Fit of the ODE SEIRDB-type Ebola model for the outbreak in Guinea in 2014. \textbf{d)} 
    Fits of the ODE SECIR-type model for SARS-CoV-2 related ICU cases in Germany for national (top) and district (bottom) level. \textbf{e)} Fits of the SECIR-type model for SARS-CoV-2 related ICU cases in Spain for national (top) and provinces (bottom) level. \textbf{f)} Evolution of infected cases for five different global and local intervention strategies for Germany. \textbf{g)} Evolution of continued NPIs and DynamicNPIs for three randomly selected districts (in color for Flensburg, in gray for Aachen and Heilbronn). \textbf{h)} Evolution of infected, severe, and critical cases for DynamicNPIs with optimal control. \textbf{i)} Summary of the five intervention strategies with respect to the average number of realized contacts and outcomes for infected, severe, and critical cases. Shape files for Germany using geodata `Verwaltungsgebiete 1:250 000 (VG250)'' from BKG (2026) dl-de/by-2-0, data sources: \url{https://sgx.geodatenzentrum.de/web_public/gdz/datenquellen/datenquellen_vg_nuts.pdf}. Shape files for Spanish administrative units CC-BY 4.0 \url{ign.es}.}
    \label{fig:scaling_metapop_and_dynamic}
    \vspace*{-0.7cm}
\end{figure}

As infectious disease dynamics vary for different diseases and are often heterogeneous on a spatial scale, we assess MEmilio's capabilities to generate, simulate, and calibrate models for different diseases and across multiple spatial scales. In particular, we investigate applications of influenza and Ebola as well as highly resolved epidemic data on SARS-CoV-2. To do so, we select three of MEmilio's pre-implemented PBMs: an SDE-based SIRS-type model for seasonal influenza allowing for seasons of varying severity, an ODE-based SEIRDB model for Ebola based on Legrand et al.~\cite{Legrand2007}, and an ODE-based SECIR-type model for the early phase of SARS-CoV-2 (\cref{fig:scaling_metapop_and_dynamic}a).

First, we evaluate the integration of MEmilio’s Python interface with Particle Markov chain Monte Carlo (PMCMC) methods \cite{Andrieu2010} for non-spatial applications to influenza and Ebola using PBMs (see Supplementary Information Section~D.1 for details). For influenza, we fit an SDE-based SIRS PBM to reported cases in Germany \cite{Grippedaten} using the pseudo-marginal Metropolis–Hastings (PMMH) algorithm. We allow for variable seasonal lengths and peak intensities, showing that three distinct seasons of influenza can be calibrated. Fluctuating reporting, in particular during holiday seasons, leads to broader uncertainty, with seasonal peak periods of the first two seasons falling in the 90~\% credibility interval (CI). The reported numbers of the summer periods as well as the broader infection dynamics of the last year falling mostly within the 50~\% CI (\cref{fig:scaling_metapop_and_dynamic}b). For Ebola, we calibrate an ODE-based SEIRDB PBM following  Legrand et al.~\cite{Legrand2007} to the 2014 Guinea outbreak~\cite{Eboladaten}, explicitly accounting for infectious and postmortem transmission. Reported case counts largely fall within the 90~\% CI and are close to the posterior median (\cref{fig:scaling_metapop_and_dynamic}c), indicating that the overall epidemic dynamics are well reproduced. Together, these results demonstrate that MEmilio’s Python interface supports likelihood-free exact Bayesian inference using PMCMC methods effectively.

Second, we test MEmilio's capabilities of generating PBMs and MPMs of different spatial resolutions and fitting the resulting models with the amortized Bayesian inference framework BayesFlow~\cite{bayesflow_2023_software}, leveraging state-of-the-art inference methods~\cite{arruda2025diffusionSBI}. Exploiting MEmilio’s resolution-agnostic model specification, we instantiate SARS-CoV-2 models for Germany at national and district (400 districts) level as well as for Spain at national and provinces (47 mainland provinces) level. Infection spread is modeled using a SECIR-type model, extending the classical SEIR formulation to represent presymptomatic transmission as well as hospitalization and critical cases requiring intensive care, and calibration targeted reported COVID-19 ICU occupancy (see Supplementary Information Section~D.2 for details).

The analysis of the calibration results reveals that the national-level models yield comparatively narrow uncertainty bands (\cref{fig:scaling_metapop_and_dynamic}d,e) but necessarily enforce homogeneous prevalence across entire countries, averaging out local deviations. At the finer spatial resolution, calibration across 400 German districts and 47 Spanish provinces also closely matches reported ICU occupancy, demonstrating that fine-grained spatial models can be effectively used when disease dynamics are driven by regional hotspots. 

By leveraging detailed spatial information, MEmilio captures regional variation in transmission dynamics, contact structures, and intervention effects, resulting in more realistic epidemic simulations despite broader uncertainty bands. This trade-off between precision and realism is fundamental: increased uncertainty at higher spatial resolution reflects genuine epidemiological heterogeneity and unresolved local influences rather than model deficiencies, providing decision makers with a more faithful representation of plausible epidemic trajectories.

\subsection*{Tailored local intervention schemes that avoid untargeted strictness}

To assess locally tailored NPIs that reduce epidemiological burden while avoiding unnecessary restrictions, we require tools that can represent policies varying across regions and over time. MEmilio supports systematic assessment of a broad range of such NPI schemes, including policies that vary across regions or dynamically adapt over time. In this section, we assess automatic NPI activation once user-defined epidemiological thresholds are exceeded and MEmilio’s integration with optimal control to balance intervention strictness against epidemiological objectives.

To study the relative effectiveness of different intervention schemes, we consider the district level model of the SARS-CoV-2 spread in Germany. We select a representative snapshot of disease dynamics based on the root mean square deviation (\cref{fig:scaling_metapop_and_dynamic}d) and perform a scenario analysis by simulating 60-days under five alternative strategies: i)~no NPIs, ii)~continuation of the calibrated NPI strictness, iii)~implementation of stricter NPIs, iv)~DynamicNPIs on district level, with predefined strictness, and v)~DynamicNPIs combined with optimal control (\cref{fig:scaling_metapop_and_dynamic}f–h). For more details, see Supplementary Information Section~E. Uncertainty from calibration was excluded by fixing the initial setting across all scenarios; thus, results should not be interpreted as forecasts, but rather as a comparative assessment of the \textit{relative} performance of different intervention strategies.

MEmilio's simulations illustrate clear differences in how the respective intervention mechanisms shape epidemic outcomes. In the absence of NPIs, approximately 70~\% of the population becomes infected, with more than 223\,000 severe and 39\,500 critical cases. Continuing the fitted NPIs reduces infections, severe cases, and critical cases by roughly a factor of three. Imposing stricter NPIs further lowers the number of critical cases and reduces the average number of daily contacts from 2.82 to 1.80. In contrast, our threshold-triggered DynamicNPIs applied for 14 days once incidences of 250 and 1\,000 symptomatic cases are exceeded achieve a comparable reduction in infections, severe, and critical cases while allowing \MK{31}~\% more contacts. When optimizing for minimal NPI strictness under the terminal constraint that total infections on January~21,~2021 remain below \MK{500}\,000, DynamicNPIs combined with optimal control reduce critical cases by further \MK{20}~\% while \MK{even} permitting \MK{35}~\% more contacts than the stricter NPI scenario.

Taken together, these results demonstrate that MEmilio provides a flexible and effective framework for implementing and evaluating targeted, local intervention strategies that can substantially reduce epidemiological burden while limiting the societal costs associated with untargeted, country-wide responses. Importantly, the same analysis pipeline can be applied without modification in the presence of parameter uncertainty, allowing the systematic evaluation of threshold-based interventions under calibrated uncertainty ensembles.

\subsection*{Reducing runtime and resource demands with hybrid modeling}

\begin{figure}[!h]
    \centering
    \includegraphics[width=1\linewidth]{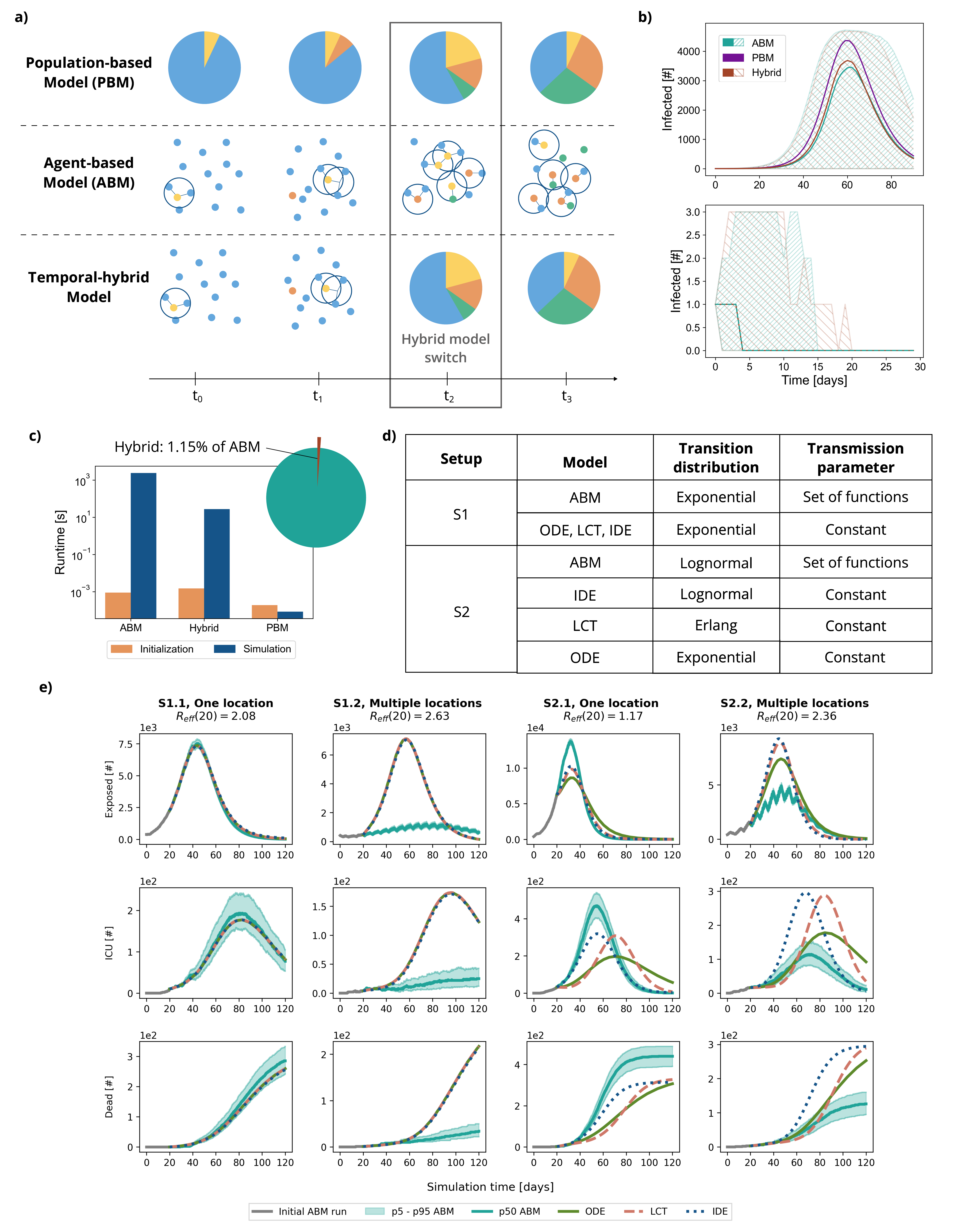}
    \caption{\textbf{PBM and ABM in comparison with temporal-hybrid ABM-PBM model and inter-model comparison with different state transition assumptions and contact structures.}\linebreak \textbf{a)} Schematic visualization of the temporal-hybrid model for a SEIR model with colors as in~\cref{fig:different_models_and_questions}. \textbf{b)}~Median and CI through 5th and 95th percentiles for 1\,000 runs of all three models for the whole simulated time frame (top) and median and 90~\%~CI of the simulations with virus extinction (bottom). \textbf{c)}~Initialization and simulation runtime for all three models, with the pie chart showing the percentage of the temporal-hybrid model's simulation runtime relative to the ABM's simulation runtime. \textbf{d)} Settings used for the inter-model comparison of ABM, ODE, LCT, IDE. \textbf{e)} Trajectories of the simulated number of Exposed, ICU, and Dead for ABM, ODE, LCT, IDE for exponential (S1) and lognormal (S2) transition distributions using one (S1.1, S2.1) or multiple (S1.2, S2.2) locations per type in the ABM. For the ABM, median and 90~\%~CI across 100 runs are shown.}
    \label{fig:hybrid_and_modelcomp}
\end{figure}

Through their increased computational cost, additional constraints can come into play when exploring a particular research question with ABMs. To test MEmilio's capabilities of simulating disease dynamics with efficient hybrid models that retain the accuracy of ABMs in highly stochastic regimes, we compare a temporal-hybrid ABM-PBM model against its pure ABM counterpart. All models use a SECIR-type disease progression and the temporal-hybrid model switches dynamically from the diffusion-based ABM to the ODE-based PBM once the predefined threshold of three infected individuals per population of 10\,000 is reached (\cref{fig:hybrid_and_modelcomp}a). For more details, see Supplementary Information, Section~F.1. 

In simulations with only one exposed individual in a total population of 10\,000, approximately 20~\% of all ABM simulations result in virus extinction. While also using PBM structures, the temporal-hybrid model is able to capture both the median number of total infected and the ABM percentiles with reasonable approximation error compared to the pure PBM(\cref{fig:hybrid_and_modelcomp}b, top). Moreover, the temporal-hybrid model reliably captures scenarios of virus extinction (\cref{fig:hybrid_and_modelcomp}b, bottom), a feature that is particularly important in local outbreaks of novel or re-emerging respiratory diseases. To quantify the computational speedup, we measured the runtime of all models. While the hybrid model requires slightly more time for initialization, it reduces the simulation runtime of the ABM by almost 99~\% (\cref{fig:hybrid_and_modelcomp}c). 

Summarized, our findings show that MEmilio's temporal-hybrid model combines the advantages of ABMs and PBMs. The hybrid model is able to capture overall disease dynamics and cases of extinction with small additional model error at a fraction of the ABM runtime and computational cost.

\subsection*{Model ensembles and quantification of model-specific assumptions}

As all models are mathematical simplifications, their \MK{outcomes depend on explicit and implicit modeling assumptions. For} example, exponentially distributed transition times, as used in most compartmental models, can lead to substantially premature or delayed predictions of epidemic peak timings~\cite{ploetzke_lct_2025}. \MK{We therefore use MEmilio’s harmonized model implementations as a cross-fidelity sensitivity-analysis framework. By varying individual assumptions while keeping the remaining epidemic setting aligned, inexpensive PBMs can be used to screen which structural choices materially affect the quantities of interest and where the additional resolution of an ABM becomes important. Such analyses can also guide surrogate construction by identifying the parameters, structural features, and epidemic regimes that need to be varied in the training data of a surrogate for a higher-fidelity model.}

\MK{To demonstrate this functionality, we consider two setups. In Setup~S1, state transition times are exponentially distributed for all models, whereas in Setup~S2, the ABM, LCT-, and IDE-based models use the specific distributions given in \cref{fig:hybrid_and_modelcomp}d. In addition, we assess the effect of replacing one large, homogeneously mixing location of each type (Home, School, Work, Event, and Shop) by multiple smaller locations in the ABM.} To generate a common starting point for all models, we create synthetic initial data by simulating one outbreak trajectory with the ABM for the first 20 days. For more details, see Supplementary Information Section~F.2.

Given Setup~S1 with one location per location type (S1.1), we first observe that all considered models can be aligned under the same simplifying assumptions, giving almost identical epidemic dynamics with respect to the number of Exposed, ICU, and deceased cases (\cref{fig:hybrid_and_modelcomp}e S1.1). Employing more realistic distributions in S2.1 results in higher and sharper peaks in the number of Exposed and ICU cases than when using exponential distributions as in the ODE-based model (\cref{fig:hybrid_and_modelcomp}e S2.1). Although the IDE- and LCT-based models approximate the median ABM peak number of Exposed closer than the ODE-based model, they both yield 25~\% reduced peaks. A similar pattern is observed for the ICU occupancy: median peak ICU cases in the ABM are approximated closer by LCT- and IDE-based models with the IDE-based model also aligning with the peak time point of the ABM. In contrast, the LCT- and ODE-based models exhibit a delay in the ICU peak of approximately 20 days (\cref{fig:hybrid_and_modelcomp}e S2.1). Replacing one location per type in the ABM by many smaller locations per type yields smaller and more heterogeneous contact networks of individuals. As a result, we observe substantially reduced transmission events and consequently less ICU admissions and deaths (\cref{fig:hybrid_and_modelcomp}e S1.2, S2.2), a property which is not captured with homogeneous mixing as modeled by the PBMs. While this effect is qualitatively known from graph theory and earlier network-based infectious disease models~\cite{keeling_networks_2005,pastor-satorras_epidemic_2015}, the comparison allows quantifying these differences appropriately.

Combining a broad range of models with harmonized interfaces and uniform model descriptions, MEmilio \MK{enables systematic cross-fidelity sensitivity analyses within a single framework. Inexpensive PBMs and MPMs can be used to screen large parameter spaces and alternative structural assumptions, while selected regimes can subsequently be evaluated with more detailed ABMs. The resulting comparisons quantify when coarse-grained models provide adequate approximations and when individual contact structures, nonexponential transition times, or other high-fidelity features materially alter the outcomes. They can therefore inform both the selection of an appropriate model resolution and the design of training and validation scenarios for surrogate models of computationally expensive simulations.}

\subsection*{Efficient and scalable implementations for swift simulations} 

\begin{figure}[!h]
    \centering
    \includegraphics[width=1\linewidth]{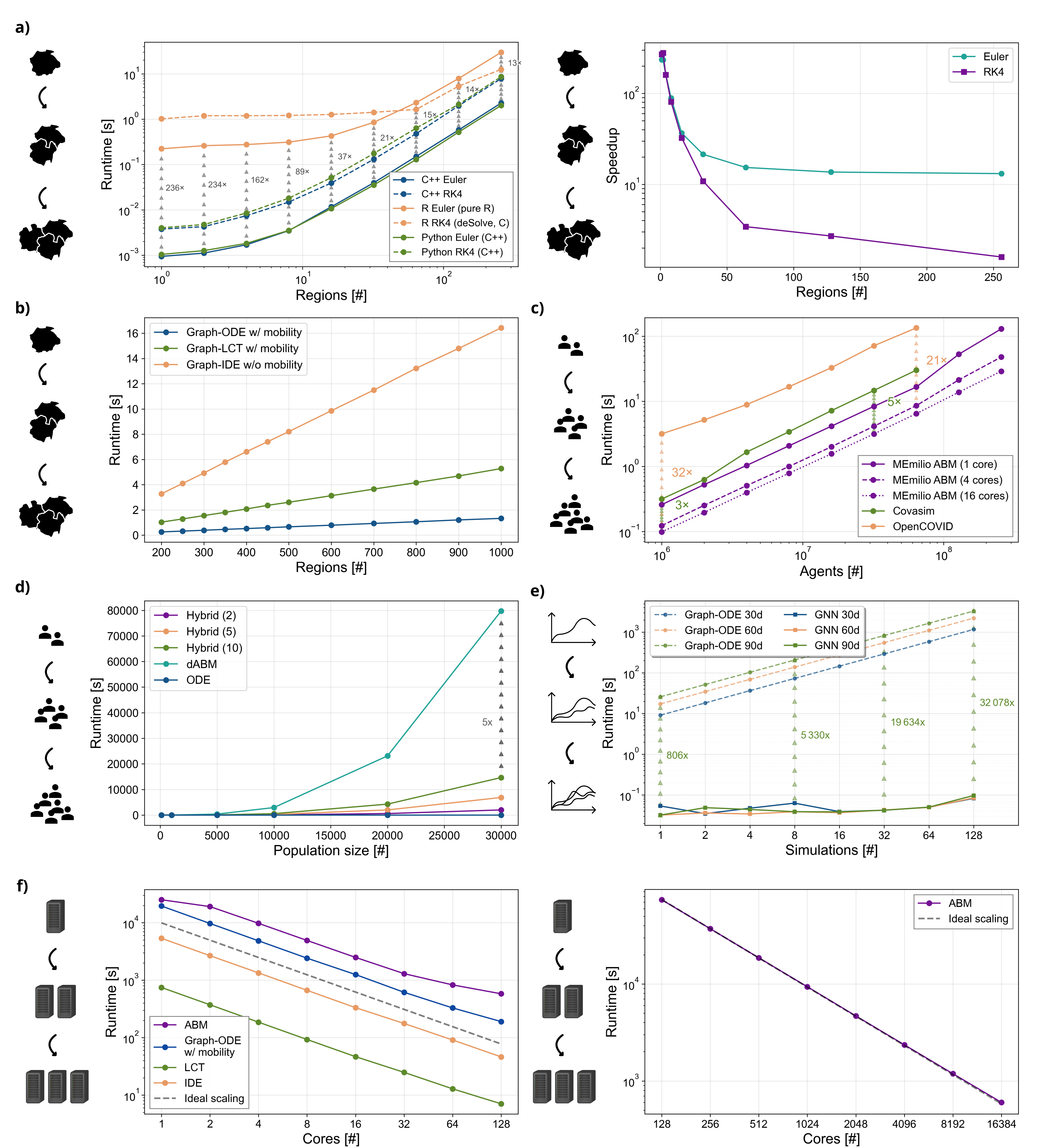}
    \caption{\textbf{Performance and computational scalability of MEmilio's models with respect to regions, population sizes, and the number of deployed CPU cores.} \textbf{a)} Performance comparison in runtime (left) and speedup (right) of MEmilio's ODE-based SEIR-type MPM (C++ in blue, Python in green) against an R implementation. \textbf{b)} Runtime scaling of MEmilio's MPMs with respect to the number of modeled regions. \textbf{c)} Runtime scaling of MEmilio's mobility-based ABM with a comparison to other frameworks. \textbf{d)} Comparison of different temporal-hybrid models to the corresponding diffusion-based ABM and ODE-based PBM in linear (left) and log (right) scale. \textbf{e)} Runtime scaling for a MPM with 400 districts and different prediction horizons (30, 60, and 90 days) compared to a correspondingly trained GNN. \textbf{f)} Strong scaling of the ensemble-run functionality for different PBMs, MPMs, and ABMs of MEmilio.}
    \label{fig:scaling}
\end{figure}

As simulation and calibration can be time-critical, efficient and scalable implementations are essential for epidemic modeling. To demonstrate MEmilio's capabilities of running large-scale models in narrow time frames and executing a large number of parallelized model runs, we assess MEmilio's efficiency as well as the scaling performance with respect to the number of regions, the population size, the number of simulations, and the number of available computing resources. For more details, see the Supplementary Information, Sections~F.1, G,~H,~I, and~J.

The assessment of MEmilio's efficiency with PBMs and MPMs reveals MEmilio's strength in building on a highly efficient C++ backend, simulating 500 days of a single PBM in the order of $10^{-3}$ seconds and up to 100 regions in a second or less (\cref{fig:scaling}a, left). In particular, the computational overhead of using MEmilio's Python interface is negligible, thus enabling efficient model executions from a user-friendly interface, e.g., through Python and Jupyter Notebooks. While Euler and RK4 methods are chosen here for a standardized comparison, additional speedups are expected when using MEmilio's default adaptive solvers. Using a manually implemented Euler solver, both of MEmilio's interfaces outperform an R-based implementation (as often used for epidemiological modeling; see~\cref{fig:model_comparisons}b) of the MPM by one or two orders of magnitude (\cref{fig:scaling}a, right). The use of the precompiled, C-based \textit{deSolve} integrator in R yields a method that comes closer to MEmilio's performance at 256 regions, but is substantially outperformed by MEmilio for modest numbers of regions.

Beyond its efficiency, we assess MEmilio's scalability with respect to larger numbers of regions for different local model types (ODE, LCT, or IDE). Regardless of the specific PBM assigned to the graph nodes, the graph-based MPM scales modestly with respect to the number of modeled regions (\cref{fig:scaling}b). Due to its substantially more complex form, the IDE-based model~\cite{Wendler2025IDE} shows a runtime that is about three times higher than that of the LCT-based model and even approximately 15 times higher than the runtime of the ODE-based model. However, all considered models complete within several seconds for up to $1\,000$ regions. 

In order to allow scalable research studies on an individual-based level, we assess the scalability of MEmilio's ABM with realistic mobility patterns. First, we observe optimal linear scaling with respect to the number of modeled individuals (agents), consistent with reports of comparable state-of-the-art ABMs such as Covasim~\cite{kerr_covasim_2021} and OpenCOVID~\cite{shattock_impact_2022}. Excluding differing features of the considered models, the sequential execution of one time step in MEmilio's ABM is 1.2-1.8 (Covasim) and 8-12 (OpenCOVID) times faster than one time step in Covasim and OpenCOVID for simulations with 1 million agents (\cref{fig:scaling}c). MEmilio’s ABM scales to substantially larger populations, successfully simulating 128 and 256 million agents, whereas Covasim and OpenCOVID did not complete runs beyond 64 million agents in our benchmark setup. We also see that we can further speedup MEmilio's ABM by using its intra-model parallelization (\cref{fig:scaling}c). 

Beyond computational optimization and direct population-size scaling in ABMs, we assess the scaling of MEmilio's previously introduced temporal-hybrid model with respect to the population size for different threshold values. In our tests, the hybridization significantly accelerates simulations with up to 30\,000 modeled individuals; expecting higher acceleration with increasing population sizes. The hybrid model's runtime depends on the chosen switching condition and has the same scaling behavior as the diffusion-based ABM (dABM) on which it is based (\cref{fig:scaling}d). While a switch upon exceeding a threshold of three infected agents in a population of 10\,000 reduced the runtime of the corresponding dABM by almost 99~\% (\cref{fig:hybrid_and_modelcomp}c), a threshold upon exceeding 10~\% of infected agents in a population of 30\,000 still results in a speedup of 5.0 (\cref{fig:scaling}d) and thresholds of five or two percent realize speedups of 11.6 and 39.8, respectively.

Finally, we assess MEmilio's scalability with respect to the number of individual simulations to either address the uncertainty in a selected setup or to consider large numbers of alternative scenarios. For MEmilio's surrogate modeling package designed to train and evaluate GNNs with spatio-temporal disease dynamics, we find that up to 128 simulations on the German mobility network with 400 districts can be executed on-the-fly within a fraction of a second (0.1~s) (\cref{fig:scaling}e). Since the corresponding simulation of the graph-based ODE MPM (Graph-ODE) takes approximately 10 seconds for one simulation, scaling linearly with the number of simulations, the speedup of the GNN ranges from 806 to 32\,078 (\cref{fig:scaling}e). Using MEmilio's ensemble-run functionality that leverages distributed memory parallelism to start multiple independent simulations in parallel (and post-process their results), we observe (almost) ideal strong scaling for ABM, Graph-ODE-based MPM, LCT- and IDE-based PBM (\cref{fig:scaling}f, left) -- apart from slightly suboptimal scaling of the ABM between one and two cores as well as 32 and 128 cores. On one hand, execution times of the MPM in \cref{fig:scaling}e can be reduced substantially by leveraging HPC resources, while, on the other hand, the pre-trained GNN allows for the provision of tens or hundred of simulations on a laptop or in a web interfaces without additional need for HPC connection. Furthermore, with up to 16\,384 cores on the JURECA supercomputer, we can obtain 16\,384 simulation results of an ABM with two million agents in approximately 17 minutes (\cref{fig:scaling}f, right).

We conclude that MEmilio's focus on efficiency and parallelization ensures minimized simulation time and enables highly parallelized execution. MEmilio's distributed memory parallelism is particularly well suited for standalone ensemble runs and offline generation of training data for GNNs~\cite{schmidt_gnn_2025} and neural network–based inference methods~\cite{bayesflow_2023_software}.

\section*{Discussion}

To overcome the existing fragmentation of mathematical modeling software, we developed MEmilio as a computationally scalable, multi-model and multi-scale epidemic modeling framework with a user-friendly interface. MEmilio allows the study of newly unfolding or ongoing epidemic dynamics in a narrow time frame to address the need for fast and efficient tools that can support public health decision-making. MEmilio's different model types provide computational scalability with respect to population sizes in ABMs, the number of modeled regions in MPMs, and, generally, the computational resources deployed to the epidemic modeling task. MEmilio's unique combination of individual-, population- and meta-population-based models together with hybridization approaches, high performance computing techniques, and graph neural networks in a single framework provides a novel and holistic approach to epidemiological modeling.

The building blocks of the presented MEmilio framework have already been used for a multitude of projects. This includes the study of NPI lifting during vaccination campaigns~\cite{koslow_appropriate_2022} for SARS-CoV-2, commuter-based testing~\cite{kuhn_regional_2022}, the adaption of human behavior~\cite{ZUNKER2025116782}, the explicit consideration of infection during travel~\cite{zunker_novel_2024}, coupling individual-based models with waste water dynamics~\cite{SCHMID2025100836,Bicker_coupled_2025}, advanced testing strategies on multi-dimensional outcomes~\cite{KERKMANN2025110269}, and additional applications in, e.g.,~\cite{kuhn_assessment_2021,Wendler2025IDE,korf_effect_2025,Kerkouche2026}. Furthermore, MEmilio includes published models for RSV~\cite{WEBER200195}, influenza~\cite{weidemann_is_2017}, or Ebola~\cite{Legrand2007}, and allows the rapid and straightforward integration of additional models for, e.g., Dengue~\cite{RODRIGUES20141,NDII2015157}, using MEmilio's SBML interface.

\MK{We showed that MEmilio’s uniform model descriptions allow demographic stratification and the integration of spatio-temporal dynamics with only limited adaptations of the model implementation. Beyond simplifying the construction of individual models, this harmonization enables models of different fidelity to be evaluated under matched epidemiological assumptions. The comparisons presented here can therefore be interpreted as a structured sensitivity analysis: by varying transition-time distributions and contact structures, we identify assumptions for which coarse-grained PBMs closely approximate the ABM and regimes in which individual-level structure leads to substantial differences.

This cross-fidelity perspective also provides a basis for more targeted surrogate development. Computationally inexpensive PBMs and MPMs can be used to screen broad parameter ranges, structural assumptions, and intervention scenarios, thereby identifying the regimes that should be sampled more densely with an expensive ABM. The resulting ABM simulations could subsequently be used to train and validate a surrogate specifically in the regions where high-fidelity effects are relevant. In the present study, we demonstrate the constituent capabilities of this workflow through harmonized inter-model comparisons and through a GNN surrogate of a graph-based MPM. Extending the surrogate target directly to the ABM is an important next step rather than a result claimed here.

More generally, multi-model ensembles can improve forecast accuracy relative to individual models~\cite{reich_accuracy_2019,cramer_evaluation_2022}. Within MEmilio, such ensembles can be realized to make model-dependent uncertainty and sensitivity to structural assumptions more transparent. Together with modular software testing and continuous integration, this supports the robustness and traceability of model implementations and their comparisons.}

\MK{Optimizing massively parallel simulations remains a persistent challenge. While most often, parallelization of MPMs is highly effective over a large number of ensemble simulations, balancing load and communication overhead within graph-based MPMs on heterogeneous architectures is non-trivial and a subject of our own active research. Consequently, future developments will target fully integrated numerical solvers for distributed memory systems. Furthermore, to make MEmilio sustainable and ready for most-recent supercomputing infrastructures, we are progressing towards hybrid or fully GPU-based~\cite{gerstein_hpc} model implementations.} 

As demonstrated, MEmilio can be efficiently coupled with established frameworks for state-of-the-art neural network–based inference methods~\cite{bayesflow_2023_software}, Markov chain Monte Carlo~\cite{pypesto}, and Approximate Bayesian Computation~\cite{schalte_pyabc_2022} to allow for a broad range of applications in parameter inference and model calibration. Furthermore, MEmilio has been successfully coupled with established packages for uncertainty quantification and sensitivity analysis~\cite{razavi_new_2016,herman_salib_2017}. Being aware of the inherent complexity of a framework with diverse model types and implementations, we have extensively reworked MEmilio's documentation and provide more than 70 code examples. In addition, we are actively working on the provision of advanced tutorials to further minimize entry barriers for new users and maximize MEmilio's applicability.

\section*{Materials and Methods}

In the following methodological section, we briefly describe MEmilio's models that are depicted in the central part of~\cref{fig:MEmilio_overview} and that have been used to generate the results of~\cref{fig:different_models_and_questions,fig:scaling_metapop_and_dynamic,fig:hybrid_and_modelcomp,fig:scaling}. More technical background on amortized Bayesian inference, mathematical solvers, parallelization, or optimal control as used for this manuscript is given in the Supplementary Information.
  
\section*{Population-based models}

MEmilio implements PBMs based on ordinary, stochastic, and integro-differential equations. PBMs provide a computationally efficient representation of infectious disease dynamics. In MEmilio, they can be used standalone or as the basis for MEmilio's more general MPMs, described in the section "Metapopulation models".

\subsection*{Models based on ordinary differential equations}

MEmilio's PBMs in the form of ODEs, commonly referred to as compartmental models~\cite{brauer_mathematical_2019}, have been used in~\cref{fig:different_models_and_questions,fig:scaling_metapop_and_dynamic,fig:hybrid_and_modelcomp,fig:scaling}. In these models, the population is partitioned into a finite set of compartments representing epidemiological states such as susceptible, exposed, infectious, or recovered, optionally stratified by sociodemographic characteristics.

MEmilio's ODEs are formulated in a \textit{flow-based formulation}~\cite{schmieding_flows} in contrast to standard approaches that typically integrate the net change of compartment sizes directly~\cite{brauer_mathematical_2019}; also see Zunker et al.~\cite{zunker2026lagrangian}. Through this specification, MEmilio directly computes particular in- and outflows of disease states as well as the total compartmental values. Let $N_C \in \mathbb{N}$ denote the number of compartments and let 
\begin{align*}
y : \mathbb{R} \to \mathbb{R}_{\geq 0}^{N_C}
\end{align*}
be the state vector, where $y_i(t)$ denotes the population size of compartment $i$ at time $t$. Transitions between compartments are described by flow functions
\begin{align*}
\sigma_{j,i} : \mathbb{R}_{\geq 0}^{N_C} \times \mathbb{R} \to \mathbb{R}_{\geq 0},
\end{align*}
where $\sigma_{j,i}(y(t),t)$ represents the rate at which individuals move from compartment $j$ to compartment $i$ at time $t$. Transitions that are not part of the model are assigned a rate of zero and omitted from the implementation. The model dynamics are then defined in terms of cumulative flows
\begin{align*}
\mathcal{F}_{j,i} : \mathbb{R} \to \mathbb{R}_{\geq 0},
\end{align*}
which represent the total number of individuals that have transitioned from compartment $j$ to compartment $i$ up to time $t$. These cumulative flows evolve according to
\begin{align}
\label{eq:flow_ode_form}
\begin{aligned}
\frac{\mathrm{d}\mathcal{F}_{j,i}(t)}{\mathrm{d}t} &= \sigma_{j,i}(y(t),t), \quad \forall j \neq i, \\
\mathcal{F}_{j,i}(t_0) &= 0.
\end{aligned}
\end{align}

The compartment sizes are reconstructed from the initial state $y(t_0) = y^0$ by accounting for the accumulated inflows and outflows,
\begin{align}
\label{eq:compartment_update_flows}
y_i(t) = y_i^0 + \sum_{j \neq i} \mathcal{F}_{j,i}(t) - \sum_{k \neq i} \mathcal{F}_{i,k}(t).
\end{align}

Population changes not arising from internal disease dynamics, such as births or deaths, are handled by introducing additional source or sink compartments without epidemiological interpretation.

MEmilio also supports direct formulation of compartment derivatives; however, the flow-based representation is used by default and throughout this work. Demographic stratification, for example by age group, is supported for most models by initializing the system with a specified number of groups, resulting in replicated compartment structures with group-specific parameters.

In this work, we use several of MEmilio's ODE-based PBMs. In~\cref{fig:scaling_metapop_and_dynamic}c, a SEIRDB-type PBM is fitted to Ebola case data for Guinea (see Supplementary Information, Section~D.1.2 for details). In~\cref{fig:scaling_metapop_and_dynamic}d,e (national level), a SECIR-type model is fitted against case data of SARS-CoV-2 in Germany and Spain (see Supplementary Information, Section~D.2 for details). The SECIR-type PBM is further used in~\cref{fig:hybrid_and_modelcomp}a-c (see Supplementary Information, Section~F.1 for details), in~\cref{fig:hybrid_and_modelcomp}d,e (see Supplementary Information, Section~F.2 for details), and in~\cref{fig:scaling}d (see Supplementary Information, Section~F.1 for details). MEmilio furthermore provides ODE-based PBMs allowing for multi-layer immunity with implicit variant change~\cite{koslow_appropriate_2022}, waning immunity with different waning paces for protection against symptomatic and severe infections~\cite{zunker_novel_2024}, a SEIRV-type influenza model~\cite{weidemann_is_2017}, and an RSV model with four layers of infection and recovery~\cite{WEBER200195}. Furthermore, by leveraging the SBML import interface, MEmilio can import a large range of models described in the SBML standard. If not stated otherwise, ODE-based initial value problems are solved numerically using an adaptive Runge-Kutta-scheme (see Supplementary Information B.1.1 for details).

\subsubsection*{Models based on the Linear Chain Trick}
\label{sec:LCT}

MEmilio uses the LCT to overcome exponentially distributed transition times that are implicitly used in simple ODE-based models and which are considered unrealistic from an epidemiological point of view~\cite{donofrio_mixed_2004,lloyd_realistic_2001,wearing_appropriate_2005}.

Using the LCT, MEmilio divides the compartments that are neither initial nor absorbing states into subcompartments. Based on the property that the sum of independent and identically distributed exponential random variables is Erlang-distributed, we obtain Erlang-distributed transitions through the linear coupling of the subcompartments. In MEmilio's LCT-based models, the single formula for one compartment of a classical ODE model is substituted by a system of formulas, with one for each subcompartment. More explicitly, the equation for compartment $i$ as in~\cref{eq:compartment_update_flows} is substituted by $N_{C,i}$ formulas where $N_{C,i}$ represents the number of subcompartments into which compartment $i$ is divided. We then have
\begin{align}
\begin{aligned}
    y_{i_1}(t) &= y_{i_1}^0 + \sum_{j \neq {i_1}} \mathcal{F}_{j,{i_1}}(t) - \sum_{k \neq {i_1}} \mathcal{F}_{{i_1},k}(t),\\
    &\,\,\,\vdots\\
    y_{i_{N_{C,i}}}(t) &= y_{i_{N_{C,i}}}^0 + \sum_{\quad j \neq {i_{N_{C,i}}}} \mathcal{F}_{j,{i_{N_{C,i}}}}(t) \quad- \sum_{\quad k \neq {i_{N_{C,i}}}} \mathcal{F}_{{i_{N_{C,i}}},k}(t).\\
\end{aligned}
\end{align}

If the variance of the state transition distribution is known, it can be used to compute the number of subcompartments to select the most appropriate Erlang distribution. While most implementations hard-code the numbers of subcompartments used~\cite{hurtado_building_2021}, MEmilio allows to set the number of subcompartments for each compartment and each group independently and the corresponding model is only created upon compile-time, leading to flexible and performant LCT-based models. In Pl\"otzke et al.~\cite{ploetzke_lct_2025}, we showed that a generalization of a model with eight compartments into one with more than 100 subcompartments only increases simulation time from 0.012 to 0.015 seconds. 

In this work, we use a SECIR-type LCT-based PBM in~\cref{fig:hybrid_and_modelcomp}d,e (see Supplementary Information, Section~F.2 for details) and in~\cref{fig:scaling}f (see Supplementary Information, Section~I.2 for details). If not stated otherwise, LCT-based initial value problems are solved numerically using an adaptive Runge-Kutta-scheme (see Supplementary Information B.1.1 for details)

\subsection*{Models based on integro-differential equations}
\label{sec:IDE}
As a generalization of LCT-based models, MEmilio provides highly flexible models based on IDEs. These models offer the possibility to consider arbitrary, user-defined or data-driven transition distributions. While IDE-based models are sometimes also referred to as age-of-infection models, MEmilio's integral formulations allow for the inclusion of any type of memory regarding disease states and as any other type of PBM, MEmilio allows to stratify the IDE-based model by demographic features such as age.

Similar to ODE-based models and as a matter of harmonization, MEmilio formulates IDE-based models using the flows from one compartment to another, i.e., computing the compartment values based on the flows. Through their memory properties, IDE-based models do not share the Markov property and, in contrast to ODE-based models, the flows $\sigma_{j,i}$ of IDE-based models do not only depend on the current state of the system but on the history of infection dynamics. Hence, infection dynamics cannot be described by a system of ODEs as in~\cref{eq:flow_ode_form} and instead, we obtain a system of Volterra integral equations as follows
\begin{align}
    \begin{aligned}
        \sigma(t) &= g(y(t),t)\,\int_{-\infty}^t f(\sigma(\tau), \tau, t)\,\d\tau, 
    \end{aligned}
\end{align}
where $\sigma$ is the vector containing all flows $\sigma_{j,i}$. The function $f$ defines the integrand of the Volterra integral equation system and $g$ is a function that may introduce nonlinearities to the system. Note that the flows may contain derivatives of the compartments which needs to be accounted for in the numerical approach. As initial conditions we require the flows $\sigma$ to be given on a sufficiently large time interval before the simulation start. 

In this work, we use a SECIR-type IDE-based PBM in~\cref{fig:hybrid_and_modelcomp}d,e (see Supplementary Information, Section~F.2 for details) and in~\cref{fig:scaling}f (see Supplementary Information, Section~I.2 for details). As default Runge-Kutta solvers cannot be applied to the initival value problems with IDE formulations, MEmilio implements a customized numerical scheme~\cite{Wendler2025IDE, messina_non-standard_2022} (see Supplementary Information, Section~B.1.2 for details). 

\subsection*{Models based on stochastic differential equations}
\label{sec:SDE}

MEmilio allows SDE-based modeling to account for random fluctuations that might substantially influence disease dynamics, especially at low population sizes. MEmilio's SDE-based PBMs generalize the ODE description by incorporating stochastic diffusion terms that can be interpreted as diffusion approximations of the underlying continuous-time Markov chain while retaining the familiar compartmental structure.

While several equivalent formulations exist for describing the diffusion component of stochastic differential equation (SDE) compartmental models~\cite{allen2008introduction}, MEmilio's approach links the diffusion terms directly to the flows between compartments. In this construction, each transition rate is accompanied by its own stochastic source so that random fluctuations are tied to the same mechanisms driving the deterministic dynamics. This flow-based perspective has been formalized by Allen et al.~\cite{allen2008construction} and provides a natural bridge between the deterministic ODE formulation and its stochastic counterpart. More precisely, we allow each cumulative flow to evolve not only according to its deterministic rate but also with a noise term that scales with the flow intensity, i.e., we replace the ODEs in~\cref{eq:flow_ode_form} by SDEs of the form
\begin{align}
\begin{aligned} \label{eq:flow_sde_form}
\d \mathcal{F}_{j,i}(t) &= \sigma_{j,i}(y(t), t)\,\d t \;+\; \sqrt{\sigma_{j,i}(y(t), t)}\,\d W_{j,i}(t), \quad \forall j \neq i,
\end{aligned}
\end{align}
where $\sigma_{j,i}$ corresponds to the flow function defined in the ODE section and $W_{j,i}(t)$ are independent standard Wiener processes. The compartment sizes are then updated as in~\cref{eq:compartment_update_flows} via the net balance of flows.

As random fluctuations in the SDE may drive the system outside the intended and non-negative region when populations approach biological boundaries, potentially leading to degenerate behavior, MEmilio post-processes the compartment values to mitigate non-negativity while preserving population sizes.

In addition to the basic SDE-SIR model, MEmilio by default provides an SDE model with waning immunity and a model with two variants or pathogen strains.

In this work, we use a SIRS-type SDE-based PBM in~\cref{fig:scaling_metapop_and_dynamic}b to model influenza cases in Germany (see Supplementary Information, Section~D1.1 for details). As in the case of IDE-based models, standard Runge-Kutta solvers cannot be applied to SDE-based models and MEmilio, thus, implements an Euler-Maruyama scheme (see Supplementary Information, Section~B.1.3 for details).

\section*{Metapopulation models}

MEmilio leverages its PBMs, as visualized in~\cref{fig:MEmilio_overview}, to model outbreak dynamics in spatially heterogeneous settings by adding an additional mobility layer to create an MPM. More precisely, the MPM divides the total population $N$ into $N_P$ distinct subpopulations (distributed over $N_P$ spatial \textit{patches}), indexed by $k \in \{1, \dots, N_P\}$.
The total population is then defined by
\begin{align*}
    N(t) = \sum_{k=1}^{N_P} N_k(t),\qquad N_k(t)=\sum_{i=1}^{N_C} y_i^k(t),
\end{align*}
where $y^k(t)=(y_1^k(t),\dots,y_{N_C}^k(t))^\top$ denotes the local compartment vector of patch $k$. The global state is the stacked vector $y(t)=(y^1(t),\dots,y^{N_P}(t))\in \mathbb{R}_{\ge 0}^{N_P\times N_C}$. Through different mobility concepts as presented in the next subsections, the patches are coupled via the exchange of individuals and commuting groups.
In the following, we distinguish between graph-based models, which are constructed from many local PBMs in an explicit graph structure, and models that connect individual PBMs through additional equations, e.g., ODEs.

\subsection*{Graph-based metapopulation model}
In the graph-based MPM, the spatial structure is represented by a directed generalized graph $\mathcal{G} = (\mathcal{V},\mathcal{E})$ where $\mathcal{V} = \{1,\dots,N_P\}$ denotes the set of nodes, each corresponding to a local subpopulation. The generalized graph uses multiple edges $E_l=(i,j)$ with $i,j \in \{1,\dots,N_P\}$, $l\in\mathbb{N}$ that implement different edge properties. More precisely, we use one particular edge for the mobility of a specific subgroup that is composed of all local individuals with a selected disease state and set of demographic features. The graph is implemented in a modular fashion, allowing for any local model to be used as a node. The graph-based MPM advances local disease dynamics within each node independently and then applies the mobility scheme along each graph edges.
There are three different mobility approaches implemented for the graph-based MPM that realize the exchange of individuals via the graph edges: a \emph{deterministic instant mobility}, a \emph{deterministic detailed mobility} which extends the previous one, and a \emph{stochastic mobility}.

\paragraph*{Deterministic instant mobility}
The deterministic instant mobility scheme, introduced in K\"uhn et al.~\cite{kuhn_assessment_2021}, aims at representing daily commuting activity. The number of individuals exchanged per day along a graph edge is given by commuting weights that represent the relative share of the population commuting between the corresponding regions (nodes). For Germany, these weights have been derived from data provided by the German Federal Employment Agency~\cite{bmas_pendlerverflechtungen_2020}. Based on the commuting weights, a half-day exchange of commuters takes place with optional reductions in mobility for symptomatically infected individuals. Additionally, commuter testing is supported, enabling the detection of infectious commuters during travel~\cite{kuhn_regional_2022}.
Since the model's resolution is at population level, individual infection trajectories cannot be tracked explicitly and infection states of commuters upon return have to be estimated. This is done by applying a one-step integration scheme separately for the commuters, while treating the other people in the same region as contact persons.
The instant mobility scheme assumes instantaneous travel and neglects travel times. As all regions exchange commuters synchronously, the scheme has ideal properties for parallelization.

The graph-based MPM, leveraging an ODE-based SECIR-type PBM with deterministic instant mobility is used in~\cref{fig:scaling_metapop_and_dynamic}d-i (see Supplementary Information, Sections~D.2 and E for details) and in~\cref{fig:scaling}b,f (see Supplementary Information, Section~I for details). The graph-based MPM is furthermore used with deterministic instant mobility, leveraging ODE-, LCT- and IDE-based SECIR-type PBMs, in~\cref{fig:scaling}b,e  (see Supplementary Information, Sections~H~and~I for details).

\paragraph*{Deterministic detailed mobility}
The deterministic detailed mobility scheme was introduced in Zunker et al.~\cite{zunker_novel_2024} and extends the instant approach by explicitly accounting for travel times and transit interactions. Each region is assigned a local mobility model, which can be flexibly parameterized to represent internal mobility patterns. 
For two regions $k$ and $l$, represented as polygons, a straight line connecting their centroids is constructed. If $k$ and $l$ are not adjacent regions, the line traverses other regions. The total predefined travel time between $k$ and $l$ is distributed uniformly across the regions intersected by this line. 
Commuters move sequentially through the corresponding local mobility models as they traverse the regions along their trip, allowing for interactions with other commuters during transit.

\paragraph*{Stochastic mobility}
\label{sec:stochastic_mobility}
The stochastic mobility scheme implements transitions between two regions as stochastic jumps given by transition rates that depend on the origin and destination node as well as on the infection state. The stochastic jumps are realized through Poisson processes and simulated using a temporal Gillespie algorithm~\cite{vestergaard_temporal_2015}. The magnitude of the rates determines the frequency of exchange via a given edge, i.e., higher rates correspond to more individuals changing from one node to another in a given time period.

\subsection*{Stochastic Metapopulation Model}
The Stochastic Metapopulation Model (SMM) describes both infection state adoptions and spatial transitions between disjoint regions as stochastic jumps governed by inhomogeneous Poisson processes. Hence, the evolution of the system state is a Markov process given by the master equation~\eqref{eq:master_equation} with $L$ the location change operator determined by spatial transition rates and $G$ the infection state transition operator as defined in the dABM section. A detailed description of the SMM can be found in Winkelmann et al.~\cite{winkelmann_mathematical_2021}. Since the whole population in one subregion has the same position, the system only changes when an infection state transition or a spatial transition takes place. Consequently, Gillespie's direct method~\cite{gillespie_approximate_2001} is used for simulation. 

\subsection*{ODE-integrated metapopulation model}

MEmilio also allows to realize an MPM, and thus spatial heterogeneity, in a large and coupled system of ODEs. This mobility pattern can be regarded as an alternative to the deterministic instant mobility approach in the graph-based MPM, with the key distinction that mobility is modeled implicitly. Instead of explicit population exchanges, the equations include the impact of infectious individuals in other patches, which is proportional to the commuting weights between two patches, through the force of infection. For a patch $k$, the flow rates between two compartments $j$ and $i$ are given by 
\begin{align*}
    \sigma_{j, i}^k: \mathbb{R}^{N_P\times N_C}\times\mathbb{R}\longrightarrow\mathbb{R}
\end{align*}
resulting in the system 
\begin{align*}
    \begin{aligned}
    \frac{\d \mathcal{F}^k_{j,i}(t)}{\d t} &= \sigma_{j,i}^k(y(t), t), \quad \forall j,i \in \{1, \dots, N_C\}, \, j \neq i, \forall k \in \{1, \dots, N_P\} \\
    \mathcal{F}^k_{j,i}(t_0) &= 0, \quad t_0\in\mathbb{R}.
\end{aligned}
\end{align*}

In this work, the ODE-integrated MPM is used with a SEIR-type structure in~\cref{fig:scaling}a (see Supplementary Information, Section~G for details).

\section*{Agent-based models}
Agent-based models (ABMs) are powerful approaches for infectious disease dynamics, addressing many of the limitations of PBMs and MPMs through a direct modeling of individuals and their interactions. ABMs define a representative, synthetic population of agents that influence and infect each other through contact patterns and individual interactions, always acting and reacting according to specified rules and their common environment. Through this individual resolution, ABMs can naturally include any heterogeneity in the population of interest. MEmilio provides two ABMs. MEmilio's main ABM allows for activity modeling and data-driven mobility rules and is denoted as \textit{mobility-} or \textit{activity-based ABM}. Additionally, MEmilio provides a mathematically rigorously described ABM based on stochastic diffusion and drift processes.

\subsection*{Mobility-based agent-based model}

MEmilio's mobility-based ABM models the movement of individual agents between manually defined locations. Instead of modeling an explicit contact network that propagates transmissions and allows infections, agents execute individual activities such as going to work or to a social event, i.e., explicitly changing to locations where they might catch an infection from other infectious agents at the same location. This approach offers a high level of flexibility when setting up an ABM. While in contact network settings, location-specific NPIs, such as venue closures, capacity restrictions, or testing scheme obligations, are modeled implicitly, the mobility-based model realizes an explicit implementation of these NPIs. 

Infections are implemented to take into account the viral load in the host, which is responsible for the amount of virus shed. Viral load dynamics are modeled as a time-dependent function, with, by default, rapid exponential growth followed by gradual decline. The transmission between agents at shared locations depends mainly on three components: age-stratified contact matrices, e.g., derived from empirical survey data, the viral shed exposure rate calculated from infectious agents present at that location weighted by contact rates between age groups, and a transmission parameter establishing a linear relationship between exposure and infection probability. Object-oriented implementations of vaccinations and face masks as well as isolation, general contact reduction, and seasonality can be used to reduce infection probability.

Aside from continuous viral load dynamics and for the purpose of harmonized model descriptions, the model also follows a discrete symptomatic state structure that is in line with our PBMs and which can be leveraged to influence the agents' behavior and movement structure, e.g., that symptomatically or severely infected individuals move less. Susceptible agents can become infected and either progress through a nonsymptomatic course of the disease and recover, or through a symptomatic course with potential escalation to severe or critical state and death. The progression can be set to depend on the agent's age and virus type or variant, if multiple types or variants are modeled.

Movement between locations can be realized in one of two ways: either through mobility rules that represent typical daily routines with random elements, or through individual data-driven trips. In both cases, a base set of infection-related movement rules is defined. This set handles transfer of severely and critically infected agents to the hospital or the ICU, and deceased agents are removed from the model. Restrictions to movement can be set up through the aforementioned interventions and testing strategies. The model also allows user-defined configurations for agent behavior, including compliance with mask mandates, testing strategies, and quarantine or isolation protocols. Through customization options, users obtain a high flexibility to simulate a wide range of intervention scenarios.

The model then progresses in discrete time steps, alternating between agent movement and simulating infection transmissions during these time steps, which are by default set to one hour. A pseudo-code is given as a general overview in Algorithm~\ref{code:sim_abm}
and more technical details can be found in Kerkmann et al.~\cite{KERKMANN2025110269}.

\begin{algorithm}
\caption{\textbf{Discrete mobility-based agent-based model}}
\label{code:sim_abm}
$t \leftarrow t_0\in\mathbb{R}$ \\
\While{$t  \leq t_{\max}$ } {
    \For{each location}{
        \textbf{Compute normalized exposure rate from infected agents at the location}    
    }
    \For{each agent}{
        \textbf{Execute agent's interactions, possibly spreading infections}
    }
    \For{each agent}{
        \textbf{Perform individual movement according to mobility rules or trips}
        }
    $t \leftarrow t+\Delta t $  
}
\end{algorithm}

To store information for millions of agents, MEmilio uses customizable \textit{loggers} to write out data points during a simulation. Predefined loggers for tracking basic movement and infection data are provided, and custom loggers can be used to define any data items and granularity to be stored in memory while the simulation is running.

In this work, the mobility-based ABM is used in~\cref{fig:hybrid_and_modelcomp}d,e (see Supplementary Information, Section~F.2 for details) and in~\cref{fig:scaling}c,f (see Supplementary Information, Section~J for details).

\subsection*{Diffusion- and drift-based agent-based model}
The diffusion- and drift-based ABM (dABM) is a mathematically motivated ABM that complements the state-of-the-art mobility-based ABM. The dABM uses a Markov Process to simulate disease dynamics and realizes agents with two features: an agent's position $x$ on a domain $\Omega \subset \mathbb{R}^2$ and its infection state $z$. The Markov Process is determined by the master equation
\begin{align} \label{eq:master_equation} 
    \partial_t p(X,Z;t) = G p(X,Z;t) + L p(X,Z;t). 
\end{align}
The operator $G$ defines the infection state adoptions, thus only acting on $Z$, the vector of all agents' infection states. Infection state adoptions are stochastic jumps modeled with independent inhomogeneous Poisson processes whose rates depend on the current system state. The operator $L$ defines location changes, i.e., movement of agents, and acts only on $X$, the vector of all agents' positions. Movement of agents is modeled with independent diffusion processes of the form
\begin{align*}
    \d x(t) = - \nabla F(x(t); t) \d t +  \omega(x(t); t)\d W,
\end{align*}
with $F(t): \Omega \to \mathbb{R}$ potentials on the domain and $\omega(t): \Omega \to \Omega$ noise functions. The Poisson processes that determine the infection state adoptions are given by rate functions that are dependent on the system state $\left(X,Z\right)$ and the current time $t$. These rates can be interpreted as the number of adoptions of a given type, e.g., from susceptible to exposed state, that happen in one unit-time step. The rates determine the time point of the next infection state adoption. In addition to system changes through infection state adoptions taking place, the system changes through time-dependent movement of agents. A time discretization for integrating the movement equation until the time point of the next infection state adoption is used, and the corresponding simulation procedure is realized by a temporal Gillespie algorithm~\cite{vestergaard_temporal_2015}.

In this work, the diffusion- and drift-based ABM is used in~\cref{fig:hybrid_and_modelcomp}a-c and in~\cref{fig:scaling}d (see Supplementary Information, Section~F.1 for details).

\section*{Hybrid agent-population and agent-metapopulation models}
\label{sec:Hybrid}

MEmilio also provides hybridized approaches, from which a temporal-hybrid model, as first introduced in Bicker et al.~\cite{bicker_hybrid_2025}, is used in this work. The idea of the temporal hybridization is based on the assumption that with lower case numbers, individual stochastic events have a substantially higher impact, while with higher case numbers, single stochastic simulation trajectories converge toward averaged outcomes. Hence, MEmilio uses PBMs and MPMs as suitable approximations for ABM simulations. The temporal-hybrid model in MEmilio combines two models and a switching condition. It continuously evaluates whether a transition between models is needed. Whenever a switch is triggered, the active model state is converted into a state of the target model through a dedicated conversion routine. These routines can be defined for arbitrary model pairs via template specialization. At present, converters are pre-implemented for the SMM and the dABM, as well as for the ODE-based SECIR-type model and the dABM. 

In this work, a SECIR-type temporal-hybrid ABM-PBM is used in~\cref{fig:hybrid_and_modelcomp}b,c, and~\cref{fig:scaling}d (see the Supplementary Information, Section~F.1 for details on setups and conversion functions).

\section*{Machine learning-based surrogate models}
An increased level of detail in modeling is almost always at the expense of the execution time of the used model. Also, despite efficient implementation, complex MPMs or ABMs can often not be simulated without cluster or HPC resources. In particular in time-critical applications, such as epidemic modeling, surrogate modeling approaches can become essential -- as they enable rapid and just-in-time evaluation of many potential scenarios.

In MEmilio, we provide surrogate modeling approaches based on several machine learning methods. These surrogate models are trained on the output of the aforementioned PBMs and MPMs and can then be used as replacement of the complex model. For the representation of complex MPMs, MEmilio provides a surrogate training structure based on GNNs. These models are able to learn spatio-temporal dynamics with connectivity patterns among regions while allowing for the integration of demographic features and NPI strictness.
While the models themselves operate as black boxes, they can be considered as intermediate between white and black boxes as they are trained on ensemble simulations integrating expert-derived mechanisms. In Schmidt et al.~\cite{schmidt_gnn_2025}, we already demonstrated that surrogate models can maintain a high level of accuracy. 

In this work, MEmilio's GNN surrogate model is used in~\cref{fig:scaling}e (see Supplementary Information, Section~H for details).

\section*{Acknowledgements}
The authors would like to thank Jan Kleinert, Annette Lutz, Wadim Koslow for early contributions to the MEmilio software and Uwe Naumann (RWTH Aachen University, i12 Software and Tools for Computational Engineering) for providing code for automatic differentiation. The authors gratefully acknowledge computing time on the supercomputer JURECA~\cite{julich_jureca_2021} at Forschungszentrum Jülich under grant HPC4EPI. 

This work was supported by the Initiative and Networking Fund of the Helmholtz Association (grant agreement number KA1-Co-08, Project LOKI-Pandemics) and by the German Federal Ministry for Digital and Transport under grant agreement FKZ19F2211A (Project PANDEMOS). It was furthermore supported by the German Federal Ministry of Education and Research and the German Federal Ministry of Research, Technology and Space under grant agreement 031L0297B (Project INSIDe), 031L0319A and 031L0319B (Project AIMS), 031L0325A (Project TwinChain), 031L0324E (Project EPISERVE), and the Deutsche Forschungsgemeinschaft (DFG, German Research Foundation) (grant agreement 528702961). TwinChain and EPISERVE are part of the Modeling Network for Severe Infectious Diseases (MONID). This study was furthermore performed as part of the Helmholtz School for Data Science in Life, Earth and Energy (HDS-LEE) and received funding from the Helmholtz Association of German Research Centres. Additionally, it was supported by the European Union via the ERC grant INTEGRATE, grant agreement number 101126146, and under Germany’s Excellence Strategy by the Deutsche Forschungsgemeinschaft (DFG, German Research Foundation) (EXC 2047—390685813, EXC 2151—390873048, and 524747443), the University of Bonn via the Schlegel Professorship of J.H. This research was partially funded by the Lower Saxony Ministry of Science and Culture (MWK) with funds from the Volkswagen Foundation's zukunft.niedersachsen program (grant number: ZN4257; Project CAIMed - Lower Saxony Center for Artificial Intelligence and Causal Methods in Medicine).

\section*{Data availability}
The pre-trained model weights for the GNN surrogate used in the benchmarks are available on Zenodo at \url{https://doi.org/10.5281/zenodo.18507331}. Contact patterns are fully integrated in the MEmilio software framework. Raw data for mobility matrices and reported case data is provided publicly by the data owners such as the Federal Employment Agency of Germany or the Robert Koch-Institute -- download and transformation to correct formats was done with memilio-epidata. 

\section*{Code availability}
The MEmilio software framework is publicly available on GitHub under \url{https://github.com/SciCompMod/memilio}. MEmilio's documentation is available at \url{https://memilio.readthedocs.io/}. The specific simulation code for this manuscript is available at \url{https://github.com/SciCompMod/memilio-simulations/tree/main/2026_Bicker_et_al_Memilio_paper}.

\section*{CRediT authorship contribution statement}


\noindent\textbf{Conceptualization:} Julia Bicker, Carlotta Gerstein, David Kerkmann, Sascha Korf, René Schmieding, Anna Wendler, Henrik Zunker, Maximilian Betz, Kilian Volmer, Michael Meyer-Hermann, Jan Hasenauer, Martin J. Kühn\\
\noindent\textbf{Data Curation:} Julia Bicker, Carlotta Gerstein, Sascha Korf, René Schmieding, Anna Wendler, Henrik Zunker, Maximilian Betz, Kilian Volmer, Patrick Lenz, Vincent Wieland\\
\noindent\textbf{Formal Analysis:} Julia Bicker, Carlotta Gerstein, Sascha Korf, René Schmieding, Anna Wendler, Henrik Zunker, Maximilian Betz, Kilian Volmer, Julian Litz, Martin J. Kühn\\
\noindent\textbf{Funding Acquisition:} Sebastian C. Binder, Margrit Klitz, Martin Siggel, Manuel Dahmen, Achim Basermann, Michael Meyer-Hermann, Jan Hasenauer, Martin J. Kühn\\
\noindent\textbf{Investigation:} Julia Bicker, Carlotta Gerstein, Sascha Korf, René Schmieding, Anna Wendler, Henrik Zunker, Maximilian Betz, Kilian Volmer, Julian Litz, Martin J. Kühn\\
\noindent\textbf{Methodology:} All authors\\
\noindent\textbf{Project Administration:} Michael Meyer-Hermann, Jan Hasenauer, Martin J. Kühn\\
\noindent\textbf{Resources:} Michael Meyer-Hermann, Jan Hasenauer, Martin J. Kühn\\
\noindent\textbf{Software:} Julia Bicker, Carlotta Gerstein, David Kerkmann, Sascha Korf, René Schmieding, Anna Wendler, Henrik Zunker, Daniel Abele, Maximilian Betz, Khoa Nguyen, Lena Plötzke, Kilian Volmer, Agatha Schmidt, Nils Waßmuth, Patrick Lenz, Daniel Richter, Hannah Tritzschak, Ralf Hannemann-Tamas, Julian Litz, Paul Johannssen, Marielena Borges, Annika Jungklaus, Manuel Heger, Annalena Lange, Elisabeth Kluth, Kathrin Rack, Martin Siggel, Martin J. Kühn\\
\noindent\textbf{Supervision:} Manuel Dahmen, Achim Basermann, Michael Meyer-Hermann, Jan Hasenauer, Martin J. Kühn\\
\noindent\textbf{Validation:} All authors \\
\noindent\textbf{Visualization:} Julia Bicker, Carlotta Gerstein, Sascha Korf, René Schmieding, Anna Wendler, Henrik Zunker, Maximilian Betz, Kilian Volmer, Vincent Wieland, Martin J. Kühn\\
\noindent\textbf{Writing – Original Draft:} Julia Bicker, Carlotta Gerstein, David Kerkmann, Sascha Korf, René Schmieding, Anna Wendler, Henrik Zunker, Maximilian Betz, Kilian Volmer, Martin J. Kühn\\
\noindent\textbf{Writing – Review \& Editing:} All authors

\bibliographystyle{naturemag}

\newpage
\begin{center}
    \Large{Supplementary Information}
\end{center}

\begin{itemize}
\item The MEmilio software framework with more than 70 code examples is available at: \\\url{https://github.com/SciCompMod/memilio}
\item A comprehensive list of tutorials and exercises for the MEmilio software framework is available at: \\\url{https://github.com/SciCompMod/memilio-tutorials/}
\item MEmilio's documentation with examples and explanations is available at:\\ \url{https://memilio.readthedocs.io/en/latest/}
\item MEmilio tutorials are actively developed at:\\ \url{https://github.com/SciCompMod/memilio-tutorials}
\item MEmilio's simulations (as used in this and other selected manuscripts) are available at: \\\url{https://github.com/SciCompMod/memilio-simulations}
\end{itemize}

\section{List of abbreviations}

In this section, we provide a list of abbreviations as used in the manuscript and the supplementary information.

\begin{itemize}
\item MEmilio: a high performance Modular EpideMIcs simuLatIOn software
\item ODE: ordinary differential equation
\item LCT: Linear Chain Trick
\item SDE: stochastic differential equation
\item IDE: integro-differential equation
\item PBM: population-based model
\item MPM: metapopulation model
\item ABM: agent-based model
\item dABM: diffusive agent-based model
\item GNN: graph neural network
\item RSV: respiratory syncytial virus
\item COVID-19: coronavirus disease 2019
\item SARS-CoV-2: severe acute respiratory syndrome coronavirus 2
\item ICU: intensive care unit
\item NPIs: nonpharmaceutical interventions
\item SBI: simulation-based inference
\item ABC: approximate Bayesian computation
\item SMC: sequential Monte Carlo
\item MCMC: Markov chain Monte Carlo
\item PMCMC: particle Markov chain Monte Carlo
\item PF: particle filter
\item MH: Metropolis-Hastings
\item PMMH: pseudo-marginal Metropolis-Hastings
\item CI: credibility interval
\item AD: automatic differentiation
\item IPOPT: Interior Point OPTimizer
\item MPI: message passing interface
\end{itemize}

\section{Mathematical tools and technical functionality}

In this section, we first give a brief description of the mathematical tools and technical functionality that are available with MEmilio. Precise usage as applied in this manuscript is described in Sections~D--I and in the simulation code published open-source.

\subsection{Numerical solvers and technical details}

All models in MEmilio are complemented by a \texttt{Simulation} class. MEmilio provides a standard implementation for ODE-based \texttt{CompartmentalModel}s and \texttt{FlowModel}s as well as a \texttt{GraphSimulation} that allows coupling multiple local simulations. More involved models, like the SDE-, IDE-, and agent-based models, define their own \texttt{Simulation}. All simulations have an \texttt{advance()} function that runs the simulation up to a user-defined time point, and in the case of the ABM also allows adding \texttt{Logger}s to specify which information to extract. The PBMs collect their results in a \texttt{TimeSeries}, storing the simulated \texttt{Population} by compartments over time, accessible through a \texttt{get\_result()} function.

\subsubsection{Numerical solvers for ODE-based models}
All ODE-based models use the same scheme. \texttt{Simulation::advance()} defines a function denoted \texttt{DerivFunction}, that expects a time $t$ and population $y(t)$ and returns the right-hand side of the model $f(y(t),t) = \frac{\d y(t)}{\d t}$, passing it to the numerical integrator.

The integrator (\texttt{OdeIntegrator}) is model independent; it only assumes that $\frac{\d y}{\d t}$ and $y$ have the same shape. It manages the result \texttt{TimeSeries}, adding new entries as it integrates from a given start time to a final time. The integrator supports explicit integration schemes with both fixed and adaptive time stepping, and the scheme can be set from the \texttt{Simulation}. By default, we use the Boost implementation~\cite{Ahnert_boost} of the Cash-Karp 54 Runge-Kutta method~\cite{cash_variable_1990}. The \texttt{DerivFunction} is used to evaluate derivatives at time points determined by the integration scheme to iteratively compute the population at the next time step.

\subsubsection{Numerical solvers for IDE-based models}
The implemented IDE-based model is solved using a nonstandard finite difference scheme that was proposed by Messina et al.~\cite{messina_non-standard_2022} and that was extended to a SECIR-type model with flow formulation Wendler et al.~\cite{Wendler2025IDE}. Here, we give a brief overview of the numerical scheme for an SIR model defined by 
\begin{align*}
    \frac{\d S}{\d t} &= -\sigma_{S,I},  \\
    \frac{\d I}{\ dt} &= \sigma_{S,I}-\sigma_{I,R}, \\
    \frac{\d R}{\ dt} &= \sigma_{I,R},
\end{align*}
where the flows between the infection states are given by 
\begin{align*}
    \sigma_{S,I} &= S(t)\lambda(t),\\
    \sigma_{I,R}(t) &= -\int_{-\infty}^t {\gamma_I}'(t-x) \, \sigma_{S,I}(x)\, \mathrm{d}x,
\end{align*}
and the force of infection is defined via
\begin{align*}
     \lambda(t) &= \frac{\phi(t)\rho(t)}{N} \int_{-\infty}^t \xi(t-x) \, \gamma_I(t-x) \, \sigma_{S,I}(x) \, \d x.
\end{align*}
$\phi$ is the average number of daily contacts, $\rho$ is the transmission probability on contact and $\xi$ denotes the fraction of infected individuals that are not isolated. $\gamma_I(\tau)$ determines the expected proportion of individuals that are still infected $\tau$ days after transmission.

Let $\Delta t>0$ be the fixed step size and $t_i = i\Delta t$, $i \in\mathbb{Z}$, equidistant time points. In the following, we use the notation $\widehat{x}$ for the respective discretized version of the variable $x$. The number of Susceptibles is determined by the number of Susceptibles in the previous time step as well as the force of infection from the previous time step by
\begin{align*}
    \widehat{S}(t_{i+1}) &= \frac{\widehat{S}(t_i)}{1+\Delta t \widehat{\lambda}(t_i)}.
\end{align*}
Using the relation $\frac{\d S}{\d t}=-\sigma_{S,I}$, we obtain
\begin{align*}
    \widehat{\sigma}_{S,I}(t_{i+1}) &= \widehat{S}({t_{i+1})} \widehat{\lambda}(t_i),
\end{align*}
which is a right endpoint approximation in $\widehat{S}$ and a left endpoint approximation in $\widehat{\lambda}$ where $\widehat{\lambda}(t_i)$ is known from the previous time step. 

Based on this, we compute the subsequent flow by
\begin{align*} 
    \widehat{\sigma}_{I,R}(t_{i+1}) &= - \Delta t \sum_{j=a}^i {\widehat{\gamma}_I}'(t_{i+1-j}) \widehat{\sigma}_{S,I}(t_{j+1}),
\end{align*}
where ${\widehat{\gamma}_I}'$ is a suitable approximation of $\gamma_I'$. The remaining compartments are computed via
\begin{align*} 
    \widehat{I}(t_{i+1}) &= \widehat{I}(t_i) + \Delta t \left(\widehat{\sigma}_{S,I}(t_{i+1}) - \widehat{\sigma}_{I,R}(t_{i+1})\right), \\
    \widehat{R}(t_{i+1}) &= \widehat{R}(t_i) + \Delta t \, \widehat{\sigma}_{I,R}(t_{i+1}).
\end{align*}

Finally, we compute the force of infection by 
\begin{align*}
     \widehat{\lambda}(t_{i+1}) &= \frac{\phi(t_{i+1})\rho(t_{i+1})}{N} \Delta t \sum_{j=a}^i \xi( t_{i+1-j}) \, \gamma_I(t_{i+1-j}) \, \widehat{\sigma}_{I,R}(t_{j+1}),
\end{align*}
where we approximate the integral by a rectangular rule with a right endpoint approximation in $\widehat{\sigma}_{I,R}$ and a left endpoint approximation in $\xi$ and $\gamma_I$.
It can be shown that this discretization scheme conserves essential biological properties of the continuous model~\cite{messina_non-standard_2022,Wendler2025IDE}. \MK{Current developments cover the analysis and adaptation of higher-order solvers~\cite{messina_positive_2022}, in particular making these numerical schemes also available for systems with contact change points~\cite{wendler_high_order}.}

\subsubsection{Numerical solvers for SDE-based models}
\label{subsec:sdesolver}
For the numeric approximation of SDEs, MEmilio utilizes the structural similarity between the Euler approximation scheme for ODEs and the Euler-Maruyama approximation scheme for SDEs. 

Consider the SDE
\begin{align*}
    \d X(t) = a(X,t) \d t + b(X,t)\d W,
\end{align*}
with the initial condition $X(0)=x_0$, where $W$ denotes a standard Wiener process. For a fixed step size $\Delta t > 0$, the Euler-Maruyama scheme approximates the solution on the interval $I=[0,T]$ as follows.

We partition $I$ into discretization points
\begin{align*}
    0=\tau_0<\tau_1<\ldots<\tau_N\leq T,
\end{align*}
with $\tau_{i+1} - \tau_{i} = \Delta t$ for all $i=0,\ldots,N-1.$

The recursive approximation is given by
\begin{align*}
    Y_0 &= x_0, \\
    Y_{i+1} &= Y_i + a(Y_i,\tau_i)\Delta t + b(Y_i,\tau_i)\Delta W_i,
\end{align*}
with $\Delta W_i = W(\tau_{i+1})-W(\tau_i)$, where the $\Delta W_i$ are independent and identically distributed normal random variables with mean zero and variance $\Delta t$. 
Using that $\sqrt{\Delta t}N(0,1)\sim N(0,\Delta t)$, we can rewrite the above scheme as follows
\begin{align*}
    Y_0 &= X_0, \\
    Y_{i+1} &= Y_i + a(Y_i,\tau_i)\Delta t + \frac{b(Y_i,\tau_i)}{\sqrt{\Delta t}}\Delta \overline{W}_i\Delta t,
\end{align*}
with $\Delta \overline{W}_i\sim N(0,1)$. With this formulation, MEmilio uses the Euler solver to solve the SDE. 

\subsection{Parallelization with shared and distributed memory}

Within different modules and kernel functionality, MEmilio uses different approaches of parallelization. 

For most PBMs and MPMs, MEmilio provides a sequential and a parallelized ensemble-run functionality, where the latter executes a user-defined number of runs with a specified number of MPI ranks~\cite{mpi50}. At the end of the simulation, the first rank collects all simulation outputs and orders them to compute the 5th, 25th, median, 75th and 95th percentiles such that credible intervals of 50~\% and 90~\% are returned by default. A shared memory parallelism for MPMs, using OpenMP~\cite{dagum1998openmp} to compute different graph nodes and mobility in parallel, has been used in various occasions on development branches. However, since generally, a lot of parallel simulations have to be conducted, the parallelization across different model runs is more efficient and, thus, preferred by default.

While we also provide MPI-parallelized ensemble-run functionality with the computation of summary statistics for the ABM, we furthermore use shared memory parallelism through OpenMP~\cite{dagum1998openmp} inside its heavier computation kernels. Since, at each time step, agents only interact within the same location, we allow computing interactions and movement simultaneously for each location.

\subsection{Optimal control through direct single shooting}\label{sec:optimal}

In the context of epidemiological models, optimal control~\cite{betts2010practical} provides a framework for determining optimal intervention strategies~\cite{lin2010optimal,tsay2020modeling}. In MEmilio, optimal control problems are formulated using a direct single shooting approach, which relies on control discretization and state integration, also referred to as the sequential method in the literature~\cite{sass2025obscured}. The nonlinear optimization problem is solved using the interior-point solver IPOPT~\cite{Waechter2006}.

To enable optimal control, MEmilio integrates automatic differentiation (AD) as a core component. AD allows the computation of exact derivatives of model outputs with respect to the control variables. The AD implementation in MEmilio builds on code developed by the group Software and Tools for Computational Engineering (STCE) at RWTH Aachen University~\cite{Naumann2011} and supports both forward- and reverse-mode automatic differentiation through operator overloading. In this approach, standard floating-point types used in computations are substituted with an active AD type. Arithmetic operators and relevant MEmilio components are overloaded for the active type such that derivative information is automatically propagated alongside the stored variable. MEmilio provides AD for a wide range of epidemiological ODE-based models, ranging from compartmental models to more intricate graph-based formulations.

\section{General calibration processes}

In this section, we briefly describe MEmilio's general functionality of being coupled with external tools for simulation-based inference (SBI) and particle filters (PF). Precise usage for this manuscript is described in Section~D.

\subsection{Calibration with simulation-based inference methods}\label{sec:cal_sbi}

MEmilio supports parameter inference via SBI, which is particularly well-suited for mechanistic epidemiological models where simulations can be conducted efficiently while the likelihood function may be expensive to evaluate, difficult to specify, or just analytically unavailable due to complex stochastics. In SBI, inference is performed by combining a parameter prior with simulations generated from the model, replacing explicit likelihood evaluations by learned or approximate inference procedures~\cite{arruda2025diffusionSBI}.

A simulator $f$ takes a parameter vector
${\theta} \in \Theta$
and produces model outputs
${y} = f(\theta)$,
typically time series such as compartment trajectories, incidences, or derived observables.
Given a prior distribution $p(\theta)$, SBI proceeds by repeatedly drawing parameters from the prior, simulating corresponding trajectories, and using the resulting pairs $(\theta, {y})$ to infer the posterior distribution
$p(\theta \mid {y}_{\mathrm{obs}})$
for an observed dataset ${y}_{\mathrm{obs}}$.

This modular structure separates the model from the inference engine: MEmilio provides the simulator output ${y}$, while inference is carried out by external libraries (e.g., BayesFlow~\cite{bayesflow_2023_software} and pyABC~\cite{schalte_pyabc_2022}) which need simulated data and prior samples to approximate the posterior distribution.
While MEmilio's backend can be used for offline generation of training data for neural network-based inference, the Python interface of MEmilio is designed such that its simulator can be coupled online into SBI libraries. 
The same simulator can therefore be used with multiple inference strategies:
(i) neural posterior estimation (as implemented in BayesFlow), and
(ii) approximate Bayesian computation with sequential Monte Carlo (ABC-SMC as implemented in pyABC). While the main manuscript focuses on the BayesFlow integration, the overall structure applies analogously to pyABC: in both cases, inference relies on repeated calls to the simulator and a prior specification over model parameters.

\subsection{Calibration with particle filters}
Particle Markov chain Monte Carlo (PMCMC) methods combine a PF with a Markov chain Monte Carlo (MCMC) scheme to enable Bayesian inference for nonlinear and non-Gaussian state space models, where the likelihood is intractable. The main idea is that a PF can produce an unbiased estimate of the likelihood and this estimate can then be used inside an MCMC acceptance step leaving the target density, i.e., the parameter posterior, invariant~\cite{Andrieu2010}. 

A PF is a sequential Monte Carlo (SMC) technique based on importance sampling with resampling primarily used for estimating the latent state distribution by an ensemble of simulated trajectories (particles)~\cite{Chopin2020}. The main idea is to iterate over the following three steps (i) propagate forward through the latent dynamics, (ii) weight according to agreement with the data under the observation model, and (iii) resample to discard low-weight particles and focus computational effort on particles that are more consistent with the data. Importantly for parameter inference, a PF produces as a by-product a likelihood estimate obtained by accumulating the normalized weights of each observation step. 

In this work, we leverage a so-called bootstrap filter, where the propagation step is performed by simulating directly from the model's latent dynamics rather than using a tuned proposal. The trade-off is that for informative observations or a high-dimensional model, the weights can become highly uneven and a large number of particles must be used to avoid particle degeneracy. We then further combine this with an outer Metropolis-Hastings (MH) scheme by replacing the likelihood term in the acceptance step with the PF estimate leading to a version of a pseudo-marginal Metropolis-Hastings (PMMH) algorithm~\cite{Andrieu2009}. The use of a bootstrap filter within an MH scheme yields a practically useful and easily applicable version of simulation-based exact Bayesian inference for stochastic models~\cite{Wieland2025}.

\section{Details on models and calibration for influenza, Ebola, and spatio-temporal SARS-CoV-2 use cases}

In this section, we provide details on models and calibration as used for Fig.~4 in the main manuscript.

\subsection{Calibration with particle filters for influenza and Ebola}

\subsubsection{SDE-SIRS model for influenza}

We use an SIRS model with seasonal transmission as used for influenza in Edlund et al.~\cite{Edlund2011}, using an SDE instead of an ODE formulation. The model equations are given by
\begin{align*}
    \d S &= -\beta(t) \frac{S\,I}{N}\, \d t + \frac{1}{T_R}R\, \d t - \sqrt{\beta(t)\frac{S\,I}{N}}\, \d W_{1} + \sqrt{\frac{1}{T_R}R}\, \d W_3, \\
    \d I &= \beta(t) \frac{S\,I}{N}\, \d t - \frac{1}{T_I}I\, \d t+ \sqrt{\beta(t)\frac{S\,I}{N}}\, \d W_{1} - \sqrt{\frac{1}{T_I}I}\, \d W_2, \\
    \d R &= \frac{1}{T_I}I\, \d t - \frac{1}{T_R}R\, \d t + \sqrt{\frac{1}{T_I}I}\, \d W_2 - \sqrt{\frac{1}{T_R}R}\, \d W_3.
\end{align*}
The effective transmission rate $\beta(t)$ is given by
\begin{align*}
    \beta(t)=\beta_0\left[\sum_i \left(\kappa_i+(1-\kappa_i)\frac{g(t-t_i,\sigma_i)}{g(0,\sigma_i)}\zeta(t,t_i)\right)\right],
\end{align*}
where $\beta_0$ is the baseline transmission coefficient, $(1-\kappa_i)$ is the fraction by which the transmission varies during season $i$, $t_i$ is the day of the peak transmission for season $i$, $\zeta(t,t_i)$ joins seasons with an apodization window of $\pm 30$ days and $\sigma_i$ is the standard deviation of the Gaussian probability distribution function $g(x,\sigma_i$).

The model is calibrated to weekly incidences of influenza in Germany, as provided by the Robert-Koch Institute~\cite{Grippedaten}. Summed over all age groups, we extracted the three seasons of the pre-COVID-19 period 2016-W32~(Start day: August 7th, 2016) to 2019-W31~(End day: August 5th, 2019).

As observables, we used the $I$ compartment with additive Gaussian noise, where the variance is scaled by the size of the $I$ compartment with
\begin{align*}
Y\sim\mathcal{N}(I,\sigma I).
\end{align*}

The initial state is given by a population size of $100\,000$, as data is provided on incidence per 100\,000 level. Of the population, 3280 are initially infected (first data point), 20\,000 initially recovered, and the remaining share of the population being susceptible.

We first fit a Gaussian curve to each season's data series. From these fits we extracted the values of $t_i$ and  $\sigma_i$ and treated them as fixed. The observation noise level is fixed to $\sigma=0.4$. The other parameters are then estimated using the PMMH method using 50 particles and 10\,000 iterations of the outer MH scheme. The parameters with bounds are given in~\cref{tab:SIRS_model_parameters}. 

\begin{table}[h]
    \centering
    \caption{\textbf{Parameters with bounds for SDE SIRS influenza model in Fig.~4b.}}
    \begin{tabular}{llll}
    \toprule
    \textbf{Symbol} & \textbf{Description} & \textbf{Support} & \textbf{Initial value} \\
    \midrule
    $\beta_0$ & Baseline transmission rate & [0.01, 0.5] & 0.05\\
    $T_I$ & Days being infected & $[1.0, 14.0]$ & 7.0\\
    $T_R$ & Days being immune & $[1.0, 50.0]$ & 14.0\\
    $\kappa_1$ & Seasonal influence season 1 & $[0.1, 0.5]$ & 0.2\\
    $\kappa_2$ & Seasonal influence season 2 & $[0.1, 0.5]$ & 0.2\\
    $\kappa_3$ & Seasonal influence season 3 & $[0.1, 0.5]$ & 0.2\\
    \bottomrule
    \end{tabular}
    \label{tab:SIRS_model_parameters}
\end{table}

\subsubsection{ODE-SEIRDB model for Ebola}

We use an SEIRDB model based on the SEIHDR model presented in Legrand et al.~\cite{Legrand2007}, explicitly using a dead but not-buried compartment for postmortem transmission. The formulation used in the manuscript is given by the following set of equations 
\begin{align*}
    \frac{\d S}{\d t} &= -(\rho_I I + \rho_D D ) S, \\
    \frac{\d E}{\d t} &= (\rho_I I + \rho_D D ) S -\frac{1}{T_E} E, \\
    \frac{\d I}{\d t} &= \frac{1}{T_E} E- \frac{1}{T_I} I, \\
    \frac{\d R}{\d t} &= \frac{\mu_I^R}{T_I} I, \\
    \frac{\d D}{\d t} &= \frac{1-\mu_I^R}{T_I} I - \frac{1}{T_B} D, \\
    \frac{\d B}{\d t} &= \frac{1}{T_B} D.
\end{align*}
The model is calibrated to the Ebola outbreak in Guinea in 2014~\cite{Eboladaten,plotdigitizer}. The start day is January 1st, 2014 and the last day is December 31st, 2014.

As observables, we use the flows from $E\rightarrow I$ together with a negative binomial noise model 
\begin{align*}
Y\sim NB(d,p)\quad\text{with}\quad p=X/(X+d),
\end{align*}
where $d$ is the dispersion parameter and $p$ the success probability.

As initial state, we use a population size of $1\times 10^7$, which is roughly the population size of Guinea and with 10 initially infected. All other individuals are designated as susceptible.

We fix the dispersion to $d=5.0$ and estimate all other parameters jointly using the PMMH method described above with 50 particles and 10\,000 iterations of the outer MH scheme. The parameters with bounds are given in~\cref{tab: SEIRDB model parameters}. 

\begin{table}[h]
    \centering
    \caption{\textbf{Parameters with bounds for ODE SEIRDB Ebola model in Fig.~4c.}}    
    \begin{tabular}{llll}
    \toprule
    \textbf{Symbol} & \textbf{Description} & \textbf{Support} & \textbf{Initial value} \\
    \midrule
    $\rho_I$ & Transmission probability on contact with infected & $[0.0001, 0.5]$ & 0.005\\
    $T_E$ & Days being exposed & $[1.0, 30.0]$ & 5.0\\
    $T_I$ & Days being infected & $[1.0, 50.0]$ & 4.0\\
    $\mu_I^R$ & Probability to recover & $[0.01, 0.99]$ & 0.5\\
    $\rho_D$ & Transmission probability on contact with deceased & $[0.0001, 0.5]$ & 0.025\\
    $T_B$ & Time until buried & $[1.0, 20.0]$& 2.0\\
    \bottomrule
    \end{tabular}
    \label{tab: SEIRDB model parameters}
\end{table}

\subsection{Calibration with BayesFlow for spatio-temporal SARS-CoV-2}

For early SARS-CoV-2 in autumn 2020 where neither vaccination nor waning immunity played a substantial role for disease dynamics, we use a SECIR-type model as presented in Kühn et al.~\cite{kuhn_assessment_2021}. In addition to a standard SEIR model, we use a pre-symptomatic infectious state as well as a severe (hospitalization) and a critical (ICU) state. The local model is described by the equations
\begin{align}
\label{eq:SECIR}
    \frac{\d S}{\d t} &= - \frac{S}{N-D}\, \phi\,\rho\, \Big( \xi_C\, C+ \xi_I\, I\Big), \nonumber \\
    \frac{\d E}{\d t} &= \frac{S}{N-D}\, \phi\,\rho\,\Big( \xi_C\, C+ \xi_I\, I\Big) 
    - \frac{1}{T_E}\, E, \nonumber \\
    \frac{\d C}{\d t} &=  \frac{1}{T_{E}}\,E  
    - \frac{1}{T_{C}}\, C, \nonumber \\
    \frac{\d I}{\d t} &= \frac{{\mu_{C}^{I}}}{T_{C}} \,C 
    -\frac{1}{T_{I}}\, I , \\
    \frac{\d H}{\d t} &= \frac{\mu_{I}^{H}}{T_{I}}\, I
     - \frac{1}{T_{H}}\, H, \nonumber \\
    \frac{\d U}{\d t} &= \frac{\mu_{H}^{U}}{T_{H}}\, H
     - \frac{1}{T_{U}} \,U ,\nonumber \\
    \frac{\d R}{\d t} &= \frac{{1-\mu_{C}^{I}}}{T_{C}}\, C
     + \frac{1 - \mu_{I}^{H}}{T_{I}}\,I
     + \frac{1-\mu_{H}^{U}}{T_{H}}\, H
     + \frac{1 - \mu_{U}^{D}}{T_{U}}\, U , \nonumber \\
    \frac{\d D}{\d t} &= \frac{\mu_{U}^{D}}{T_{U}}\, U. \nonumber
\end{align}

We employ amortized Bayesian inference through BayesFlow~\cite{bayesflow_2023_software} for calibration of the model. Here, a neural inference model is trained exclusively on simulated data and then applied to real observations without requiring additional model-specific optimization. Once trained, posterior inference for the observed data ${y}_{\mathrm{obs}}$ is obtained by a forward pass through the network, enabling fast inference and uncertainty quantification in downstream applications.

The BayesFlow pipeline is composed of two main neural components:
(i) a summary network which encodes the simulator output into a latent representation, and
(ii) an inference network which transforms the latent representation into a posterior approximation. These two networks can be trained jointly to provide optimal latent representation for inference~\cite{bayesflow_2020_original}.

In this work, we consider the following architectures:
\begin{itemize}
  \item Summary network: either a \textit{TimeSeriesNetwork}, which is tailored to sequential model outputs, or a \textit{FusionTransformer}, which provides a flexible transformer-based encoder for structured time series.
  \item Inference network: a \textit{FlowMatching} model, representing a state-of-the-art approach~\cite{arruda2025diffusionSBI}, capable of capturing complex posterior distributions in high-dimensional parameter spaces.
\end{itemize}

\begin{table}[!h]
    \centering
    \caption{\textbf{Prior distributions for the fitting of the ODE-based SECIR-type MPM in Fig.~4d,e.}}
    \label{tab:fitting_bf_params}
    \renewcommand{\arraystretch}{1.3}
    \begin{tabular}{lp{5.8cm}p{5.6cm}}
        \toprule
        \textbf{Symbol} & \textbf{Description} & \textbf{Distribution} \\
        \midrule
        & Factor correcting for underreported infections & $\text{Uniform}(1., 10.)$\\ 
    $\bar{r}$ & Mean contact reduction factor & $\text{Uniform}(0., 1.)$\\
    $r_i$ & State-specific contact reduction factor & $\text{TruncatedNormal}(\text{MEAN}=\bar{r}, \text{STD}=0.1)$ \\
    $T_E$ & Average time spent in Exposed state & $\text{Uniform}(1., 5.2)$\\
    $T_I$ & Average time spent in Infected state & $\text{Uniform}(4., 10.)$\\
    $T_H$ & Average time spent in Hospitalized state & $\text{Uniform}(5., 10.)$\\
    $T_U$ & Average time spent in ICU state & $\text{Uniform}(9., 17.)$\\
    $\mu_C^R$ & Proportion of Recovered per Carrier & $\text{Uniform}(0., 0.4)$\\
    $\mu_I^H$ & Proportion of Hospitalized per Infected & $\text{Uniform}(0., 0.2)$\\
    $\mu_H^U$ & Proportion of ICU Cases per Hospitalized & $\text{Uniform}(0., 0.4)$\\
    $\mu_U^D$ & Proportion of Dead per ICU & $\text{Uniform}(0., 0.4)$\\
    $\rho$ & Transmission probability on contact & $\text{Uniform}(0., 0.2)$\\
        \bottomrule
    \end{tabular}
\end{table}

\begin{table}[!h]
    \centering
    \caption{\textbf{Overview of the network details for calibration of the models for Germany and Spain in Fig. 4d,e.}}
    \label{tab:bf_network_details}
    \begin{tabular}{lllll}
        \toprule
        \textbf{Scenario} & \makecell[l]{\textbf{Contact change} \\ \textbf{time points}} & \textbf{Summary Network} & \textbf{Inference Network} & Epochs \\
        \midrule
        Germany (spatially resolved) & 15, 30, 45 & \texttt{FusionTransformer} & \texttt{FlowMatching}, $\operatorname{width}{=}512$ & 500\\
        Germany & 15, 25, 35 & \texttt{TimeSeriesNetwork} & \texttt{FlowMatching}, $\operatorname{width}{=}256$ & 300 \\
        Spain (spatially resolved) & 15, 30, 45 & \texttt{TimeSeriesNetwork} & \texttt{FlowMatching}, $\operatorname{width}{=}512$ & 500 \\
        Spain & 0, 20, 40 & \texttt{TimeSeriesNetwork} & \texttt{FlowMatching}, $\operatorname{width}{=}512$ & 500 \\
        \bottomrule
    \end{tabular}    
\end{table}

Training is performed using only simulated trajectories, without access to the observed dataset beyond its dimensionality and preprocessing conventions. After training, the resulting amortized posterior estimator $q$ can be applied to ${y}_{\mathrm{obs}}$ to generate posterior samples $\theta \sim q(\theta \mid {y}_{\mathrm{obs}})$. 
These posterior samples can then be simulated again to validate if estimated parameters fit the real data.

For the model calibration of Germany and Spain, we use the graph-based MPM with the deterministic mobility approach. The edge weights are derived from official registration and mobility data, respectively, for Germany~\cite{bmas_pendlerverflechtungen_2022} and for Spain~\cite{beneduce2025pyspainmobility}. We use reported data from October 1st, 2020 and before~\cite{robert_koch_institut_2025_17046865,robert_koch_institut_2025_17046972,ministerio_de_sanidad_datos_2023,ministerio_sanidad_covid19_es_2023} together with population data according to the respective spatial resolution to initialize the ODE-based models. The baseline contact rates are extracted from Prem et al.~\cite{prem_projecting_2017} and set to 7.95 for Germany and 12.32 for Spain. We assume symptomatic and asymptomatic individuals to be equally infectious, i.e., $\xi_C = \xi_I = 1$. The contact rates are adjusted every 10 to 20 days, representing the implementation of nonpharmaceutical interventions (NPIs) or behavioral adaptation. For the models with spatial resolution, NPIs and adaptation are always implemented on a federal state level. The strictness of the NPIs as well as the majority of epidemiological parameters are calibrated to the number of ICU cases~\cite{robert_koch_institut_2025_17046865,ministerio_de_sanidad_datos_2023} by neural parameter estimation through BayesFlow~\cite{bayesflow_2023_software}. For the model of Spain, we use version 2.0.7, while the model of Germany is calibrated using the updated version 2.0.8. We assume uniform prior distributions for the epidemiological parameters. Their bounds can be found in Table~\ref{tab:fitting_bf_params}. For the damping values, we assume that the strength of the NPI is similar across the country. Therefore, we draw a mean value from a uniform distribution and then draw the strength for every federal state from a truncated normal distribution with this mean. Observation noise is modeled additively on top of the deterministic model output using a spike-and-slab distribution~\cite{NEURIPS2022_db0eac67} with a spike of 0.4 and a slab of 0.2. \cref{tab:bf_network_details} gives an overview over the networks and details for each of the scenarios. For the summary network, we set the dropout probability to 0.1, and the summary dimension to twice the number of parameters. For the inference network, we only change the width of the backbone of the flow matching model. All other hyperparameters are set to their defaults in BayesFlow.

\section{Models and parameters for scenario analysis of NPIs}

In this section, we describe the scenario analysis of Fig.~4f-i. Building upon the calibrated graph-based MPM described in Section~D.2, we study the effects of different NPI strategies in a scenario analysis. To select a representative simulation, we use the simulation with the smallest root mean square deviation. The corresponding parameters can be found in~\cref{tab:params_npi}. Selecting the system state at day 60 (November 30th, 2020) as a starting point, we investigate the system's behavior under the following scenarios:

\begin{itemize}
    \item No NPIs: lift all NPIs after day 60.
    \item NPIs continued: continue with the calibrated NPIs at day 60.
    \item Strict NPIs: increase calibrated NPI strictness by a factor of 1.2.
    \item DynamicNPIs: introduce DynamicNPIs at day 60 for incidence thresholds of 250 and 1\,000 with a strictness of 0.\MK{63} and 0.\MK{77}, respectively. DynamicNPIs are enforced for 14 days if a threshold is exceeded and continued if the simulated numbers stay above the threshold.
    \item DynamicNPIs with optimal control: formulate an optimal control problem with the strictness of DynamicNPIs as control parameters. The resulting strictness values of 0.\MK{53} und 0.9 are obtained by constraining total infections on January 21, 2021 to \MK{500}\,000 and minimizing an objective function that penalizes NPI strictness if NPIs are active, i.e., when the corresponding incidence thresholds of 250 and 1\,000 are exceeded, scaled by regional population shares and aggregated across regions and time. For more details on optimal control, see Section~\ref{sec:optimal}. 
\end{itemize}

\begin{table}[!h]
    \centering
    \caption{\textbf{Parameters for NPI scenarios in Fig. 4f-i.}}
    \label{tab:params_npi}
    \renewcommand{\arraystretch}{1.3}
    \begin{tabular}{llp{6.5cm}}
        \toprule
        \textbf{Symbol} & \textbf{Description} & \textbf{Value} \\
        \midrule
    $T_E$ & Average time spent in Exposed state & 4.88165 \\
    $T_C$ & Average time spent in Carrier state & 0.31835 \\
    $T_I$ & Average time spent in Infected state & 5.55430 \\
    $T_H$ & Average time spent in Hospitalized state & 9.32235 \\
    $T_U$ & Average time spent in ICU state & 15.49097 \\
    $\mu_C^R$ & Proportion of Recovered per Carrier & 0.19291 \\
    $\mu_I^H$ & Proportion of Hospitalized per Infected & 0.00450 \\
    $\mu_H^U$ & Proportion of ICU Cases per Hospitalized & 0.17786 \\
    $\mu_U^D$ & Proportion of Dead per ICU & 0.15412 \\
    $\rho$ & Transmission probability on contact & 0.09050 \\
        \bottomrule
    \end{tabular}
\end{table}

\section{Models and parameters for model comparison}
In this section we present the models used for the temporal-hybrid model comparison (Fig.~5b-c, Fig.~6d) and the ones used for the ABM, ODE, LCT and IDE comparison (Fig.~5d-e).

\subsection{Temporal-hybrid model comparison and scaling}
 For the applications of the temporal-hybrid in Fig.~5b-c and Fig.~6d, the diffusive ABM (dABM) is coupled with the ODE-based SECIR-type model. The ODE-based SECIR-type model is given by~\cref{eq:SECIR}. For the dABM, the same infection states and state transitions are used, however infectious agents can transmit the virus only to agents within a given interaction radius $r\in\mathbb{R}$. Individual movement is defined by diffusion processes on the single well potential
 \begin{align*}
     F(x,y) = \frac{x^4 + y^4}{2},
 \end{align*}
enforcing agents to stochastically return to a center region and interact with each other. The temporal-hybrid model switches between two models according to a predefined condition. With $m_1(t)\in \Omega_1$ the system state of the first model at time $t$ and $m_2(t)\in\Omega_2$ the system state of the second model, respectively, the condition is defined by a function $\Gamma:\Omega_1\times\Omega_2 \rightarrow \{0,1\}$. If $\Gamma\left(m_1(t), m_2(t)\right)=1$, a model switch is triggered and the state of the currently active model has to be converted to the next used model using appropriate conversion functions.
For the applications in this paper, a threshold value $\tau$ is used and $\Gamma$ is given by
\begin{align*}
    \Gamma\left(m_1(t),0\right) = \begin{cases}
        1, \text{ if } (I_{total}(t)>\tau \lor I_{total}(t)<1)\\
        0, \text{ else }
    \end{cases}
\end{align*}
with $m_1(t)$ the system state of the dABM at time $t$ and $I_{total}(t)=E(t)+C(t)+I(t)+H(t)+U(t)$ the total number of infected agents at time $t$. For the application in Fig.~5b-c a threshold of three infected agents ($\tau=3$) is used and for Fig.~6d a threshold of $2~\%, 5~\%$ and $10~\%$ ($\tau\in\{0.02n_a, 0.05n_a, 0.1n_a\}$) is compared where $n_a$ denotes the number of agents.\\
Since the ODE-SECIR model lacks spatial resolution, the dABM state can be converted by aggregating agents to subpopulations according to their infection state. Conversely, converting the ODE-SECIR state to the dABM would require reconstruction of agents' positions and infection states which could be done via random sampling. However, in the presented scenarios, the temporal-hybrid model does not switch back once its active model is the ODE-based model. All parameters used for the simulations are given in~\cref{tab:params_hybrid}.

\begin{table}[!h]
    \centering
    \caption{\textbf{Parameters for temporal-hybrid applications in Fig. 5b-c and Fig. 6d.}}
    \label{tab:params_hybrid}
    \renewcommand{\arraystretch}{1.3}
    \begin{tabular}{lll}
        \toprule
        \textbf{Symbol} & \textbf{Description} & \textbf{Value} \\
        \midrule
        $n_{a}$ & Number of agents & \makecell[l]{$10\,000$ (Fig. 5b-c),\\
        $\{100, 1\,000, 5\,000, 10\,000, 20\,000, 30\,000\}$ (Fig. 6d)} \\
        $\mu_C^I$ & Proportion of Infected per Carrier & $0.75$\\
        $\mu_I^H$ & Proportion of Hospitalized per Infected & $0.01$\\
        $\mu_H^U$ & Proportion of ICU Cases per Hospitalized & $0.1$\\
        $\mu_U^D$ & Proportion of Dead per ICU & $0.05$\\
        $T_E$ & Average time spent in Exposed state& $4.5$\\
        $T_C$ & Average time spent in Carrier state & $3$\\
        $T_I$ & Average time spent in Infected state & $7$\\
        $T_H$ & Average time spent in Hospitalized state & $15$\\
        $T_U$ & Average time spent in ICU state & $16$\\
        $\lambda$ & Transmission rate & $0.5$\\
        $r$ & dABM interaction radius & $0.1$\\
        $\sigma$ & dABM noise parameter & $0.5$\\
        $E(t_0)$ & Initially Exposed & $1$ (Fig. 5b-c), $1~\%$ (Fig. 6d)\\
        $t_0$ & Simulation start & $0$\\
        $dt$ & (Initial) step size & $0.1$\\
        $t_{\max}$ & Simulation time frame in days & $90$ (Fig. 5b-c), $60$ (Fig. 6d)\\
        \bottomrule
    \end{tabular}
\end{table}

\subsection{Comparison of ABM and ODE-, LCT- and IDE-based models}

For the comparison of the mobility-based ABM and the ODE-, LCT- and IDE-based models, we consider a SECIR-type disease progression. For more realistic assumptions, we use age-resolved models. To simplify notation, we will describe the used models only for one age group in the following; see~\cref{tab:params_abm_ide_lct_all} and~\cref{tab:params_abm_ide_lct_scenario} for a description of all used parameters. 

The ODE-based SECIR-type model is given by~\cref{eq:SECIR}. The LCT-based SECIR-type model is given by 
\begin{alignat}{3}
\label{eq:LCT-SECIR}
    \frac{\d S}{\d t} &=-  \,\frac{S}{N-D}\,\phi\,\rho\, \Big( \xi_{C} \, C_{*}+\xi_{I}\, I_{*}\Big), &
    \nonumber\\
    \frac{\d E_1}{\d t}&= \,\frac{S}{N-D}\,\phi\,\rho\,\Big( \xi_{C} \, C_{*}+\xi_{I}\, I_{*}\Big)
            - \frac{n_{E}}{T_{E}}\, E_{1}, &
            \nonumber\\
    \frac{\d E_j}{\d t} &= \frac{n_{E}}{T_{E}}\, E_{j-1}
            - \frac{n_{E}}{T_{E}}\, E_{j}, &\text{ for }j\in\{2,\dots,n_{E}\}
            \nonumber\\[3pt]
    \frac{\d C_1}{\d t} &= \frac{n_{E}}{T_{E}}\, E_{n_{E}}
            -\frac{n_{C}}{T_{C}}\, C_{1}, &
            \nonumber\\[3pt]
    \frac{\d C_j}{\d t} &= \frac{n_{C}}{T_{C}}\, C_{j-1}
            - \frac{n_{C}}{T_{C}}\, C_{j}, &\text{ for }j\in\{2,\dots,n_{C}\}
            \nonumber\\[3pt]
    \frac{\d I_1}{\d t} &= \mu_{C}^{I}\,\frac{n_{C}}{T_{C}}\,             C_{n_{C}} 
            -\frac{n_{I}}{T_{I}}\, I_{1}, &
            \nonumber\\[3pt]
    \frac{\d I_j}{\d t}&= \frac{n_{I}}{T_{I}}\, I_{j-1}
            - \frac{n_{I}}{T_{I}}\, I_{j}, &\text{for }j\in\{2,\dots,n_{I}\}
            \\[3pt]
    \frac{\d H_1}{\d t}&= \mu_{I}^{H}\,\frac{n_{I}}{T_{I}}\,                I_{n_{I}}
            - \frac{n_{H}}{T_{H}}\,  H_{1}, &
            \nonumber\\[3pt]
    \frac{\d H_j}{\d t} &= \frac{n_{H}}{T_{H}}\, H_{j-1}
            - \frac{n_{H}}{T_{H}}\, H_{j}, &\text{ for }j\in\{2,\dots,n_{H}\}
            \nonumber\\[3pt]
    \frac{\d U_1}{\d t}&=  \mu_{H}^{U}\,\frac{n_{H}}{T_{H}} \,            H_{n_{H}}
            -\frac{n_{U}}{T_{U}}\, U_{1}, &
            \nonumber\\[3pt]
    \frac{\d U_j}{\d t} &= \frac{n_{U}}{T_{U}}\, U_{j-1}
            - \frac{n_{U}}{T_{U}}\, U_{j},& \text{ for }j\in\{2,\dots,n_{U}\}
            \nonumber\\[3pt]
    \frac{\d R}{\d t} &=  \left(1-\mu_{C}^{I}\right)\frac{n_{C}}{T_{C}}\,C_{n_{C}}
            + \left(1-\mu_{I}^{H}\right)\frac{n_{I}}{T_{I}}\, I_{n_{I}}, &
            \nonumber\\*
            &\quad+ \left(1-\mu_{H}^{U}\right)\frac{n_{H}}{T_{H}}\, H_{n_{H}}
            + \left(1-\mu_{U}^{D}\right)\frac{n_{U}}{T_{U}}\, U_{n_{U}}, &
            \nonumber\\[1pt]
   \frac{\d D}{\d t}&=  \mu_{U}^{D}\,\frac{n_{U}}{T_{U}} \, U_{n_{U}}, \nonumber &
\end{alignat}
where $Z_* = \sum_{j=1}^{n_Z} Z_j$ is the total number of individuals in compartment $Z$. 

The flow-based formulation of the IDE-based SECIR-type model is given by 
\begin{align}
\label{eq:IDE-SECIR}
    \frac{\d S}{\d t}&=-\sigma_{S,E} ,\nonumber \\
    \frac{\d E}{\d t}&=\sigma_{S,E}-\sigma_{E,C},\nonumber \\
    \frac{\d C}{\d t}&=\sigma_{E,C}-\sigma_{C,I}-\sigma_{C,R},\nonumber \\
    \frac{\d I}{\d t}&=\sigma_{C,I}- \sigma_{I,H}-\sigma_{I,R}, \\
    \frac{\d H}{\d t}&= \sigma_{I,H}-\sigma_{H,U}-\sigma_{H,R},\nonumber \\
    \frac{\d U}{\d t}&=\sigma_{H,U}-\sigma_{U,D}-\sigma_{U,R},\nonumber \\
    \frac{\d R}{\d t}&=\sigma_{C,R}+\sigma_{I,R}+\sigma_{H,R}+\sigma_{U,R},\nonumber \\
    \frac{\d D}{\d t}&=\sigma_{U,D},\nonumber
\end{align}
where the flows are defined via
\begin{align*}
\begin{aligned}
    \sigma_{S,E}(t)=& \,
        S(t)\,\lambda(t) ,
    \quad && &\sigma_{E,C}(t)=&
        - \int_{-\infty}^{t} {\gamma_{E}^{C}}'(t-x)\,\sigma_{S,E}(x)\, \d x,\\
    \sigma_{C,I}(t)=&
        -\mu_{C}^{I}\,\int_{-\infty}^{t} {\gamma_{C}^{I}}'(t-x)\,\sigma_{E,C}(x)\, \d x ,
    \quad && &\sigma_{C,R}(t)=&
        -\left(1-\mu_{C}^{I}\right)\,\int_{-\infty}^{t}{\gamma_{C}^{R}}'(t-x)\,\sigma_{E,C}(x)\, \d x ,\\
    \sigma_{I,H}(t)=&
        -\mu_{I}^{H}\,\int_{-\infty}^{t} {\gamma_{I}^{H}}'(t-x)\,\sigma_{C,I}(x)\,\d x ,
    \quad && &\sigma_{I,R}(t)=&
        -\left(1-\mu_{I}^{H}\right)\int_{-\infty}^{t} {\gamma_{I}^{R}}'(t-x)\,\sigma_{C,I}(x)\, \d x ,\\
    \sigma_{H,U}(t)=&
        -\mu_{H}^{U}\,\int_{-\infty}^{t} {\gamma_{H}^{U}}'(t-x)\,\sigma_{I,H}(x)\, \d x ,
    \quad && &\sigma_{H,R}(t)=&
        -\left(1-\mu_{H}^{U}\right)\int_{-\infty}^{t} {\gamma_{H}^{R}}'(t-x)\,\sigma_{I,H}(x)\, \d x ,\\
    \sigma_{U,D}(t)=&
        -\mu_{U}^{D}\,\int_{-\infty}^{t} {\gamma_{U}^{D}}'(t-x)\,\sigma_{H,U}(x)\, \d x ,
    \quad && &\sigma_{U,R}(t)=&
        -\left(1-\mu_{U}^{D}\right)\int_{-\infty}^{t} {\gamma_{U}^{R}}'(t-x)\,\sigma_{H,U}(x)\, \d x,
        \end{aligned}
\end{align*}
and the force of infection is given by
\begin{align*} 
\begin{aligned}
	\lambda(t) &= \frac{\phi(t)}{N-D(t)} \, \int_{-\infty}^t \xi_C\,\rho \left(\mu_C^I\,\gamma_C^I(t-x)+\left(1-\mu_C^I\right)\gamma_C^R(t-x)\right) \sigma_{E,C}(x)\\
     & \hspace{4cm}  + \xi_I\,\rho\left(\mu_I^H\,\gamma_I^H(t-x)+\left(1-\mu_I^H\right)\gamma_I^R(t-x)\right) \sigma_{C,I}(x) \ \d x. \\
\end{aligned}
\end{align*}
The function $\gamma_{Z_1}^{Z_2}$ denotes the survival function of the distribution with respect to $T_{Z_1}^{Z_2}$, i.e. the mean time spent in state $Z_1$ before transitioning to state $Z_2$. 

In setup S1, exponential infection state transition distributions are used for all models, and we additionally assume $T_{Z_1}^{Z_2}= T_{Z_1}$ for all feasible $Z_2$ for the ABM and IDE-based model. In setup S2, lognormal state transition distributions are used for the ABM and IDE-based model based on representative data for COVID-19~\cite{kerr_covasim_2021}. For the LCT-based model, the parameters of the Erlang distribution are chosen such that mean and variance are as close as possible to the corresponding lognormal distribution. The exponential distributions in the ODE-based model are chosen such that the mean values agree with those of the corresponding lognormal distributions.
For both setups, we additionally compared different location structures for the ABM. While in Setups S1.1 and S2.1 only one location per type is used, corresponding to homogeneous mixing in every location type, in Setups S1.2 and S2.2 multiple locations per type are used, see~\cref{tab:params_abm_ide_lct_scenario} for the applied distributions.

To ensure comparability between different models, we simulated a single trajectory with the ABM for 20 days to initialize all models with the same state. More precisely, we use the sample infection dynamics to initialize the ABM and IDE-based model whereas for the LCT- and the ODE-based model only the compartment sizes on day $20$ are used. The transmission probability on contact used in the ODE-, LCT- and IDE-based models is harmonized with the transmission rates of the ABM based on S1.1 and subsequently used for all other scenarios. Eventually, simulations of all models are run for 100 days. The comparison is done between $100$ ABM and one deterministic run for each PBM, all starting from day $20$. The effective reproduction number as provided on day $20$ is computed with the ODE-based model using the next generation matrix. 

\newgeometry{left=2cm,right=2cm,top=2cm,bottom=2cm}
\begin{landscape}
\renewcommand{\arraystretch}{1.3}
\begin{table}[h!]
    \centering
    \caption{\textbf{Joint parameters for all scenarios for ABM-IDE-LCT-ODE comparison application in Fig.~5d-e.}}
    \label{tab:params_abm_ide_lct_all}
    \begin{tabular}{llll}
        \toprule
        \textbf{Symbol} & \textbf{Scenario} & \textbf{Description} & \textbf{Value} \\
        \midrule
        $N_{G}$ & all & Number of age groups & $6$ \\
        $W_{age}$ & all & Age group distribution & $(0.0477, 0.0903, 0.2275, 0.3447, 0.2183, 0.0714)$\\
        $\mu_C^I$ & all & Proportion of Infected per Carrier & $(0.75, 0.75, 0.8, 0.8, 0.8, 0.8)$\\
        $\mu_I^H$ & all & Proportion of Hospitalized per Infected & $(0.0075, 0.0075, 0.019, 0.0615, 0.165, 0.225)$\\
        $\mu_H^U$ & all & Proportion of ICU cases per Hospitalized & $(0.075, 0.075, 0.075, 0.15, 0.3, 0.4)$\\
        $\mu_U^D$ & all & Proportion of Dead per ICU & $(0.05, 0.05, 0.14, 0.14, 0.4, 0.6)$\\
        $\phi$ & all & Contact matrix between age groups & Based on Prem et al.~\cite{prem_projecting_2017} and Fumanelli et al.~\cite{fumanelli_inferring_2012} \\
        $\rho$ & all & Transmission probability on contact & $$(0.0409163, 0.0351982, 0.0337711, 0.0333412, 0.0328736, 0.0300238)$$ \\
        $\xi_C$ & all & Proportion of Carrier individuals not isolated & $1$ \\
        $\xi_I$ & all & Proportion of Infected individuals not isolated & $1$ \\
        $v_P$ & all & Viral load peak for ABM & $8.1$\\
        $v_I$ & all & Viral load incline for ABM & $2$\\
        $v_D$ & all & Viral load decline for ABM & $-0.17$\\
        $(\alpha,\beta)$ & all & Viral shed parameters for ABM & $(-7,1)$\\
        $s_f$ & all & Viral shed factor for ABM & $\text{Uniform}(0, 0.6)$\\
        \bottomrule
    \end{tabular}
\end{table}
\end{landscape}
\restoregeometry

\newgeometry{left=2cm,right=2cm,top=2cm,bottom=2cm}
\begin{landscape}
\renewcommand{\arraystretch}{1.3}
\begin{table}[h!]
    \centering
    \caption{\textbf{Scenario-specific parameters for ABM-IDE-LCT-ODE comparison application in Fig.~5d-e.}}
    \label{tab:params_abm_ide_lct_scenario}
    \begin{tabular}{llll}
        \toprule
        \textbf{Symbol} & \textbf{Scenario} & \textbf{Description} & \textbf{Value} \\
        \midrule
        $T_E$ & S1 & Average time spent in Exposed state & $(4.5, 4.5, 4.5, 4.5, 4.5, 4.5)$\\
        $T_C$ & S1 & Average time spent in Carrier state & $(2.825, 2.825, 2.48, 2.48, 2.48, 2.48)$\\
        $T_I$ & S1 & Average time spent in Infected state & $(7.9895, 7.9895, 7.9734, 7.9139, 7.9139, 7.685)$\\
        $T_H$ & S1 & Average time spent in Hospitalized state & $(16.855, 16.855, 16.855, 15.61, 13.12, 11.46)$\\
        $T_U$ & S1 & Average time spent in ICU state & $(17.73, 17.73, 17.064, 17.064, 15.14, 13.66)$\\
        \midrule
        $T_E^C$ & S2 & Time from E to C & $\text{LogNormal}(\text{MEAN}=4.5,\text{STD}= 1.5$)\\
        $T_C^I$ & S2 & Time from C to I & $\text{LogNormal}(\text{MEAN}=1.1,\text{STD}= 0.9$)\\
        $T_C^R$ & S2 & Time from C to R & $\text{LogNormal}(\text{MEAN}=8.0, \text{STD}=2.0$)\\
        $T_I^H$ & S2 & Time from I to H & $\text{LogNormal}(\text{MEAN}=6.6,\text{STD}= 4.9$)\\
        $T_I^R$ & S2 & Time from I to R & $\text{LogNormal}(\text{MEAN}=8.0,\text{STD}= 2.0$)\\
        $T_H^U$ & S2 & Time from H to U & $\text{LogNormal}(\text{MEAN}=1.5,\text{STD}= 2.0$)\\
        $T_H^R$ & S2 & Time from H to R & $\text{LogNormal}(\text{MEAN}=18.1, \text{STD}=6.3$)\\
        $T_U^D$ & S2 & Time from U to D & $\text{LogNormal}(\text{MEAN}=10.7, \text{STD}=4.8$)\\
        $T_U^R$ & S2 & Time from U to R & $\text{LogNormal}(\text{MEAN}=18.1, \text{STD}=6.3$)\\
        $n_E$ & S2 & Number of subcompartments in the LCT model for E & $(9, 9, 9, 9, 9, 9)$\\
        $n_C$ & S2 & Number of subcompartments in the LCT model for C & $(5, 5, 4, 4, 4, 4)$\\
        $n_I$ & S2 & Number of subcompartments in the LCT model for I & $(16, 16, 16, 15, 15, 13)$\\
        $n_H$ & S2 & Number of subcompartments in the LCT model for H & $(8, 8, 8, 7, 6, 5)$\\
        $n_U$ & S2 & Number of subcompartments in the LCT model for U & $(8, 8, 8, 8, 7, 6)$\\
        \midrule 
        $W_{household}$ & S1.2, S2.2 & Household size distribution for the ABM & $(0.415, 0.342, 0.118, 0.091, 0.034)$\\
        $n_{workplace}$ & S1.2, S2.2 & Workplace size for the ABM & $\text{Normal}(\text{MEAN}=15, \text{STD}=25)$\\
        $n_{workplace}^{min}$ & S1.2, S2.2 & Minimum workplace size for the ABM & $5$\\
        $n_{school}$ & S1.2, S2.2 & School size for the ABM & $\text{Normal}(\text{MEAN}=42, \text{STD}=3)$\\
        $n_{school}^{min}$ & S1.2, S2.2 & Minimum school size for the ABM & $15$\\
        $n_{Event}$ & S1.2, S2.2 & Event size for the ABM & $\text{Normal}(\text{MEAN}=42, \text{STD}=941)$\\
        $n_{Event}^{min}$ & S1.2, S2.2 & Minimum event size for the ABM & $10$\\
        $n_{Shop}$ & S1.2, S2.2 & Shop size for the ABM & $\text{Normal}(\text{MEAN}=90,\text{STD}=1000)$\\
        $n_{Shop}^{min}$ & S1.2, S2.2 & Minimum shop size for the ABM & $30$\\
        \bottomrule
    \end{tabular}
\end{table}
\end{landscape}
\restoregeometry

\section{Cross-implementation performance: C++, Python, and R}
\label{sec:si_language_benchmark}

This section details the implementation specifics for the runtime comparison presented in Fig.~6a of the main manuscript. The objective of this benchmark is to quantify the computational efficiency of MEmilio's C++ backend -- accessed directly and via its Python interface -- against standard implementation patterns in the statistical computing language~R; which is often used for epidemiological modeling (see Fig.~1b of the manuscript).

\subsection{Model specification and metapopulation extension}
For this benchmark, we utilize the standard SEIR compartmental model without age stratification. The local dynamics for a population of size $N(t) = S(t) + E(t) + I(t) + R(t)$ are given by the following system of ODEs with
\begin{align*}
\begin{aligned}
    \frac{\d S}{\d t} &= -\,\lambda\,S, \\
    \frac{\d E}{\d t} &= \lambda\,S\;-\;\frac{1}{T_{E}}\,E, \\
    \frac{\d I}{\d t} &= \frac{1}{T_{E}}\,E\;-\;\frac{1}{T_{I}}\,I, \\
    \frac{\d R}{\d t} &= \frac{1}{T_{I}}\,I.
\end{aligned}
\end{align*}
Here, $T_E \in \mathbb{R}_{> 0}$ and $T_I \in \mathbb{R}_{>0} $ denote the average latency and infectious periods, respectively. The force of infection $\lambda_i$ is defined as
\begin{align*}
    \label{eq:si_lambda_benchmark}
    \lambda = \rho \, \phi \, \frac{I}{N},
\end{align*}
where $\rho \in [0,1]$ is the transmission probability and $\phi \in \mathbb{R}_{>0}$ denotes the mean daily contact rate.

To evaluate scalability, we extend the PBM to an MPM using the ODE-integrated mobility scheme as presented in the methods section. The resulting global system of ODEs has a dimension of $4 N_P$, where $N_P$ is the number of patches. The benchmark scales the number of modeled regions $N_P$ from 1 to 256. To ensure a computational load on the mobility handling, we assumed a fully connected topology where individuals can infect from infected individuals of all regions; weighted with the corresponding commuter exchange. The mobility coefficients are initialized uniformly.

\subsection{Implementations}
We compare four implementation approaches to solve the resulting system of ODEs:

\begin{enumerate}
    \item \textbf{MEmilio C++:} Direct execution of the compiled C++ library using both the explicit Euler and the Runge-Kutta-4 (RK4) integrator.
    \item \textbf{MEmilio Python (with precompiled C++ code):} Execution via MEmilio's Python bindings. This setup uses the identical C++ core for calculation but handles the initialization and function calls through Python, testing the overhead introduced by the Python interface layer.
    \item \textbf{R (Pure):} A native R implementation of the explicit Euler scheme. This represents a baseline for ad-hoc scripts as used for rapid prototyping or modeling tasks.
    \item \textbf{R (with precompiled deSolve):} An implementation using the deSolve package in R, which uses compiled C/Fortran code for the integration steps.
\end{enumerate}

\subsection{Simulation parameters}
All simulations are performed over a time horizon of $T=500$ days with a fixed step size of $\Delta t = 0.1$. The initial parameters, epidemiological rates, and population sizes used for this benchmark are identical across all implementations to ensure comparability and the values are listed in \cref{tab:example_params_r_py_cpp}. 
To mitigate the impact of execution variability, each benchmark configuration is repeated 40 times. We report the median runtime to ensure stable timings robust against outliers. The simulations are executed on a dedicated compute node equipped with an Intel Xeon Skylake CPU (Gold 6132) with 2.60 GHz and 384 GB DDR4 memory.

\begin{table}[h]
    \centering
    \caption{\textbf{Simulation parameters and initial conditions for the cross-implementation performance benchmark in Fig.~6a.} The settings are applied identically across the C++, Python, and R implementations to ensure the comparability of the runtime results.}
    \label{tab:example_params_r_py_cpp}
    \renewcommand{\arraystretch}{1.3}
    \begin{tabular}{l p{10.5cm} l}
        \toprule
        \textbf{Symbol} & \textbf{Description} & \textbf{Value} \\
        \midrule
        $N_{G}$ & Number of age groups (non-stratified) & $1$ \\
        $N_{P}$ & Number of patches (regions) scaled for the benchmark & $\{1, 2, \dots, 256\}$ \\
        $S^{(r)}(t_0)$ & Initial susceptible population per region $r$ & $10\,000$ \\
        $E^{(0)}(t_0)$ & Initial exposed individuals in region $0$ (patient zero seed) & $100$ \\
        $E^{(r)}(t_0)$ & Initial exposed individuals in regions $r > 0$ & $0$ \\
        $I^{(r)}(t_0)$ & Initial infected individuals (all regions) & $0$ \\
        $R^{(r)}(t_0)$ & Initial recovered individuals (all regions) & $0$ \\
        $\phi$ & Average number of daily contacts & $2.7$ \\
        $\rho$ & Transmission probability on contact & $0.01$ \\
        $T_E$ & Average time spent in Exposed state & $5.2$ days \\
        $T_I$ & Average time spent in Infected state  & $6.0$ days \\
        $t_{\max}$ & Simulation time frame in days & $500$ \\
        $\Delta t$ & Fixed step size for numerical integration & $0.1$ days \\
        $C_{r}$ & Retention rate (individuals staying in home patch $r$) & $0.8$ \\
        $C_{rs}$ & Mobility rate ($r \neq s$): Outflow distributed uniformly to all other patches (fully connected) & $\frac{0.2}{N_{P}-1}$ \\
        \bottomrule
    \end{tabular}
\end{table}

\section{Benchmarking Graph Neural Network surrogates against metapopulation model}
\label{sec:si_gnn_benchmark}

This section describes the setup for the performance comparison between the mechanistic graph-based MPM based on the formulation in Kühn et al.~\cite{kuhn_assessment_2021} and the Graph Neural Network (GNN) surrogate model based on Schmidt et al.~\cite{schmidt_gnn_2025}, as presented in Fig.~6e of the main manuscript. The objective is to quantify the computational speedup achieved by replacing the numerical integrator with a learned surrogate in large-scale ensemble scenarios.

\subsection{Surrogate model architecture and training}
The surrogate model is implemented using TensorFlow~\cite{tensorflow2015} and the Spektral~\cite{grattarola_graph_2021} library for graph deep learning. The architecture is adapted from Schmidt et al.~\cite{schmidt_gnn_2025} and designed to capture spatio-temporal dynamics on the graph of Germany's 400 districts ($N=400$) considering $N_G=6$ distinct age groups.

Specifically, the network consists of a GNN with seven ARMA convolutional layers (ARMAConv)~\cite{bianchi_graph_2021}, each configured with 512 channels and ReLU activation. The final layer is a dense projection layer that maps the node features to the output time-series (flattened trajectory of all compartments over the simulation horizon).  Input features include the binary connectivity matrix of the mobility network and the full initial state vector, comprising eight epidemiological compartments for each of the six age groups across all 400 districts.

To ensure the surrogate generalizes across different epidemiological regimes, it is trained on a dataset of 1\,000 distinct mechanistic simulations. These simulations are generated by varying the initial conditions, specifically the magnitude and spatial distribution of the initial infections, in a graph-based MPM. The dataset is split into training (80~\%), validation (10~\%), and test (10~\%) sets. Training is performed using the Adam optimizer with a mean squared error loss function. To ensure robust convergence and prevent overfitting, we employ early stopping based on the validation loss. Detailed information on the parameter distributions used for dataset generation can be found in Schmidt et al~\cite{schmidt_gnn_2025}.

\subsection{Benchmark setup}
We compare the runtime of both approaches for three distinct prediction horizons (30, 60, and 90 days) and varying simulation batch sizes $B \in \{1, 2, 4, \ldots, 128\}$. The simulation batch size represents the number of independent simulations computed simultaneously.

\begin{itemize}
    \item \textbf{Graph-based MPM:} The reference runtimes are derived from the execution time of the optimized C++ solver for the graph-based MPM using a SECIR model (see Eq.~\cref{eq:SECIR}) for the local nodes. Since the numerical integration of independent ensemble members is performed sequentially on the benchmark hardware, the total runtime scales linearly with the number of simulations.
    \item \textbf{GNN Surrogate:} The evaluation time is measured on a GPU-accelerated node. For each configuration, we performed 100 evaluations and reported the mean wall-clock time. Thanks to the parallelism of the GPU, the evaluation time is nearly constant for all batch sizes.
\end{itemize}

The surrogate maintains a stable mean absolute percentage error of up to $15~\%$ relative to the mechanistic ground truth across all batch sizes, confirming that the massive speedup does not come at the cost of degrading approximation quality~\cite{schmidt_gnn_2025}. 
All measurements are conducted on a compute node equipped with an Intel Xeon Scalable Processor Skylake (Silver) and four NVIDIA Tesla V100 SXM2 GPUs (16 GB VRAM). The GNN inference utilizes a single GPU, while the ODE baselines are executed on the CPU.

\section{Performance evaluation of PBMs and MPMs}

In this section, we provide the models and setups used for the performance evaluation of the PBMs~as presented in Fig.~6b,f of the manuscript.

\subsection{Scaling with respect to regions in ODE-, LCT-, and IDE-based MPM}

For the region-dependent scaling in Fig.~6b we use the graph-based SECIR-type MPM using ODE-based (see~\cref{eq:SECIR}), LCT-based (see~\cref{eq:LCT-SECIR}), and IDE-based (see~\cref{eq:IDE-SECIR}) models in the local nodes. For each model we consider $N_G=1$ age groups and performed $100$ simulations for a time frame of $30$ days and a varying number of regions $N_P=\{200, 250, 300, ..., 1\,000 \}$. The ODE- and LCT-based MPM realize mobility using the instant mobility approach with a node degree of $100$. MEmilio's IDE-based MPM is not yet able to incorporate mobility, hence its simulations are run without any exchange of individuals between regions.

\subsection{Strong scaling with respect to the number of simulations}

The strong scaling setup for the PBMs in Fig.~6f uses SECIR-type models considering $N_G=6$ age groups. The models are initialized with population and case data for the early phase of the COVID-19 pandemic in Germany~\cite{robert_koch_institut_2025_17046972} and with parameters motivated by Kühn et al.~\cite{kuhn_assessment_2021}, with an added variation of $\pm 10~\%$ to all infection parameters such that the simulated scenarios differ in each run. The graph-based MPMs using ODEs represents $N_P=400$ German counties while the simulations of IDE- and LCT-based model only use one global representation i.e., $N_P=1$. Contact patterns for the graph-based MPM and LCT-based models are based on Prem et al. and Fumanelli et al.~\cite{prem_projecting_2017,fumanelli_inferring_2012}. The mobility data for the ODE-based MPM is obtained from official data from 2020~\cite{bmas_pendlerverflechtungen_2020}.
The parameter values for all models are detailed in~\cref{tab:parameters_pbm_scaling}. We performed a total of 16\,384 simulations for the LCT- and IDE-based model each and 1\,280 simulations for the graph-based MPM, scaled over 1 to 128 cores.

All PBM and MPM simulations are performed on (up to) three Intel Xeon Skylake (Gold 6132) CPU nodes with 56 cores with 2.60 GHz and 384 GB DDR4 memory.

\newgeometry{left=2cm,right=2cm,top=2cm,bottom=2cm}

\begin{landscape}

\renewcommand{\arraystretch}{1.3}
\begin{table}[h!]
    \centering
    \caption{\textbf{Parameters for scaling applications of PBMs and MPMs in Fig.~6b,f}. Values in parentheses indicate values per age group (youngest to oldest).}
    \label{tab:params_strong_scaling_pbm}
    \begin{tabular}{llll}
        \toprule
        \textbf{Symbol} & \textbf{Model} & \textbf{Description} & \textbf{Value} \\
        \midrule
        $N_G$ & all & Number of age groups & $1$ (Fig.~6b), $6$ (Fig.~6f)\\
        \midrule
        $\xi_{C}$ & ODE, LCT & Proportion of people from the Carrier compartment who are not isolated & $1.0$\\
        $\xi_{I}$ & ODE, LCT & Proportion of people from the Infected compartment who are not isolated & $0.3$\\
        $\rho$ & ODE, LCT & Transmission probability on contact & $0.07333$ (Fig.~6b), $(0.03, 0.06, 0.06, 0.06, 0.09, 0.175)$ (Fig.~6f)\\
        $\mu^{R}_{C}$ & ODE, LCT & Proportion of Recovered per Carrier & $0.206901$ (Fig.~6b), $(0.25, 0.25, 0.2, 0.2, 0.2, 0.2)$ (Fig.~6f)\\
        $\mu^{H}_{I}$ & ODE, LCT & Proportion of Hospitalized per Infected & $0.07864$ (Fig.~6b), $(0.0075, 0.0075, 0.019, 0.0615, 0.165, 0.225)$ (Fig.~6f)\\
        $\mu^{U}_{H}$ & ODE, LCT & Proportion of ICU cases per Hospitalized & $0.17318$ (Fig.~6b), $(0.075, 0.075, 0.075, 0.15, 0.3, 0.4)$ (Fig.~6f)\\
        $\mu^{D}_{U}$ & ODE, LCT & Proportion of Dead per ICU & $0.21718$ (Fig.~6b), $(0.05, 0.05, 0.14, 0.14, 0.4, 0.6)$ (Fig.~6f)\\
        $T_{E}$ & ODE, LCT & Average time spent in Exposed state & $3.335$ (Fig.~6b), $(3.335, 3.335, 3.335, 3.335, 3.335, 3.335)$ (Fig.~6f)\\
        $T_{C}$ & ODE, LCT & Average time spent in Carrier state & $2.58916$ (Fig.~6b), $(2.74, 2.74, 2.565, 2.565, 2.565, 2.565)$ (Fig.~6f)\\
        $T_{I}$ & ODE, LCT & Average time spent in Infected state & $6.94547$ (Fig.~6b), $(7.02625, 7.02625, 7.0665, 6.9385, 6.835, 6.775)$ (Fig.~6f)\\
        $T_{H}$ & ODE, LCT & Average time spent in Hospitalized state & $7.28196$ (Fig.~6b), $(5., 5., 5.925, 7.55, 8.5, 11.)$ (Fig.~6f)\\
        $T_{U}$ & ODE, LCT & Average time spent in ICU state & $13.066$ (Fig.~6b), $(6.95, 6.95, 6.86, 17.36, 17.1, 11.6)$ (Fig.~6f)\\
        \hline
        $n_{Z}$ & LCT & Number of subcompartments per compartment & $1$ (S, R, D), $5$ (E, C, I, H, U)\\
        \midrule
        $\gamma(\tau)$ & IDE & Expected proportion of individuals who remain in the considered compartment $\tau$ days after entering & $\text{SmootherCosine}(2)$\\
        $\rho$ & IDE & Transmission probability on contact & $\text{ConstantFunction}(1)$\\
        $\xi_{C}$ & IDE & Proportion of people from the Carrier compartment who are not isolated & $\text{ConstantFunction}(1)$\\
        $\xi_{I}$ & IDE & Proportion of people from the Infected compartment who are not isolated & $\text{ConstantFunction}(1)$\\
        \bottomrule
    \end{tabular}

    \label{tab:parameters_pbm_scaling}
\end{table}
\end{landscape}
\restoregeometry

\begin{table}[!h]
    \caption{\textbf{ABM Parameters for Fig.~6c,f.}
    For the proportional parameters in the second part, values are given for the different age groups in ascending order. We denote LogNormal distributions with dependency on ({mean, std}) as parameters. Contact scaling of the baseline contact matrices are provided for the five location types: Home, School, Work, SocialEvent, and BasicsShop.}
        \renewcommand{\arraystretch}{1.3}
        \begin{tabular}{l l l l}
            \toprule
            \textbf{Symbol}                   & \textbf{Description}                                                        & \textbf{Value}             & \textbf{Source}                         \\
            \midrule
            $v_{\textrm P}$                     & Viral load peak                                                             & $8.1$                                         & ~\cite{jones_estimating_2021}               \\
            $v_{\textrm I}$                     & Viral load incline                                                          & $2$                                           & ~\cite{jones_estimating_2021}               \\
            $v_{\textrm D}$                     & Viral load decline                                                          & $-0.17$                                       & ~\cite{jones_estimating_2021}                \\
            $(\alpha,\beta)$                             & Viral shed parameters                                                        & $(-7,1)$                                           & ~\cite{jones_estimating_2021}               \\
            $s_{\textrm f}$                     & Viral shed factor                                                           & Uniform($0.00,0.28$)                            & Adopted from~\cite{ke_daily_2022}               \\
            $s$                           & Contact Matrix Scaling                                              & $(21.0, 4.8, 1.5, 14.4,3.16)$                                        &~\cite{noauthor_zeitverwendungserhebung_nodate}\\
            $\lambda$                           & Infection rate from viral shed                                              & $1.0$                                        &~Manual Adoption                      \\
            \midrule
            $\mu^{I}_{C}$                 & Proportion of Infected per Carrier                                 & $(0.5, 0.55, 0.6, 0.7, 0.83, 0.9)$          & ~\cite[Tab. 2]{kerr_covasim_2021}                       \\
            $\mu^{H}_{I}$                 & Proportion of Hospitalized per Infected              & $(0.02, 0.03, 0.04, 0.07, 0.17, 0.24)$      & ~\cite{nyberg_risk_2021}                                \\
            $\mu^{U}_{H}$                     & Proportion of ICU cases per Hospitalized                 & $(0.1, 0.11, 0.12, 0.14, 0.33, 0.62)$       & ~\cite[Tab. 2]{zali_mortality_2022}                     \\
            $\mu^{D}_{U}$                     & Proportion of Dead per ICU                                     & $(0.12, 0.13, 0.15, 0.26, 0.4, 0.48)$       & ~\cite[Tab. 2]{zali_mortality_2022}                     \\
            $T^{C}_{E}$         & Time from E to C                                         & $\text{LogNormal}(\text{MEAN}=4.5,\text{STD}= 1.5)$  & ~\cite[Tab. 1]{kerr_covasim_2021} \\
            $T_{C}^{I}$     & Time from C to I                & $\text{LogNormal}(\text{MEAN}=1.1,\text{STD}= 0.9)$  & ~\cite[Tab. 1]{kerr_covasim_2021}  \\
            $T_{C}^{R}$         & Time from C to R                        & $\text{LogNormal}(\text{MEAN}=8.0, \text{STD}=2.0)$  & ~\cite[Tab. 1]{kerr_covasim_2021}\\
            $T_{I}^{H}$ & Time from I to H               & $\text{LogNormal}(\text{MEAN}=6.6,\text{STD}= 4.9)$  & ~\cite[Tab. 1]{kerr_covasim_2021}\\
            $T_{I}^{R}$     & Time from I to R                          & $\text{LogNormal}(\text{MEAN}=8.0,\text{STD}= 2.0)$  & ~\cite[Tab. 1]{kerr_covasim_2021} \\
            $T_{H}^{U}$     & Time from H to U         & $\text{LogNormal}(\text{MEAN}=1.5,\text{STD}= 2.0)$  & ~\cite[Tab. 1]{kerr_covasim_2021} \\
            $T_{H}^{R}$     & Time from H to R                               & $\text{LogNormal}(\text{MEAN}=18.1, \text{STD}=6.3)$ & ~\cite[Tab. 1]{kerr_covasim_2021}  \\
            $T_{U}^{D}$         & Time from U to D                                                & $\text{LogNormal}(\text{MEAN}=10.7, \text{STD}=4.8)$ & ~\cite[Tab. 1]{kerr_covasim_2021}  \\
            $T_{U}^{R}$         & Time from U to R                             & $\text{LogNormal}(\text{MEAN}=18.1, \text{STD}=6.3)$ & ~\cite[Tab. 1]{kerr_covasim_2021}  \\
            \bottomrule
        \end{tabular}

    \label{tab:parameters_abm}
\end{table}

\section{Performance evaluation of the mobility-based ABM}

The large-scale ABM simulations are executed on JURECA-DC supercomputing facility~\cite{julich_jureca_2021} equipped with compute nodes of 2x AMD EPYC 7742, 2× 64 cores with 512\,GB DDR4 memory per node and where we fix the frequency to 2\,GHz.

\subsection{Comparisons with other frameworks}

Runtime scaling experiments in Fig.~6c are conducted using the ABM configured with COVID-19 disease parameters as described in~\cref{tab:parameters_abm}. The simulation is initialized with approximately 0.05~\% of agents initially infected (exposed), using a synthetic population based on German census data~\cite{destatis_population_age,noauthor_erwerbstatigenquoten_nodate,noauthor_projected_nodate,noauthor_pupils_nodate}, divided into six age groups. To isolate computational overhead from logging operations, a minimal logger configuration is employed. Each benchmark run simulated five days of epidemic dynamics resulting in 120 time steps. 

For comparison, equivalent runtime scaling experiments are performed using Covasim and OpenCOVID. Similarly to our ABM, simulations are initialized with approximately 0.05~\% of the population initially infected (exposed) and run for 120 time steps of one day. Covasim benchmarks use the hybrid population generation for Germany and OpenCOVID benchmarks the default age distribution together with the default contact network for Germany. Finally, each simulation is run for 1, 2, 4, 8, 16, 32, 64, 128, and 256 million agents. Unfortunately, we were unable to complete simulations of Covasim and OpenCOVID with 128 and 256 million agents without running out of memory.

\subsection{Strong scaling with respect to the number of simulations}

Strong scaling in Fig.~6f is conducted using a population of 2\,000\,000 agents divided into six age groups. The synthetic population is generated based on German census data~\cite{destatis_population_age,noauthor_erwerbstatigenquoten_nodate,noauthor_projected_nodate,noauthor_pupils_nodate}, and the simulation is parameterized with COVID-19 disease parameters, again shown in \ref{tab:parameters_abm}. 
Results of 1 to 128 cores are conducted on a single node while results from 128 to 16\,384 cores are conducted on 1 to 128 nodes with 128 cores each. As before, 0.05~\% of agents are initially Exposed. Disease state transitions are recorded using a logger to capture the temporal evolution of epidemiological compartments. Each simulation covers a 14-day period.

\end{document}